\documentclass[preprint,12pt]{elsarticle}

\usepackage[hang]{footmisc}
\usepackage{amssymb}
\usepackage{bm}
\usepackage{psfrag}
\usepackage{amsthm,color}

\usepackage[lined,boxed,commentsnumbered]{algorithm2e}

\usepackage{epsf,amsfonts,amsmath}

\newtheorem{rem}{Remark}
\newtheorem{define}{Definition}

\journal{Elsevier}

\begin{document}

\begin{frontmatter}

\title{Low-Rank Separated Representation Surrogates of High-Dimensional Stochastic Functions: Application in Bayesian Inference}

\author{AbdoulAhad Validi}
\ead{validiab@msu.edu}

\address{Mechanical Engineering Department, Michigan State University, East Lansing, MI, 48824, USA}

\begin{abstract}
\label{Abstract}
This study introduces a non-intrusive approach in the context of low-rank separated representation to construct a surrogate of high-dimensional stochastic functions, e.g., PDEs/ODEs, in order to decrease the computational cost of Markov Chain Monte Carlo simulations in Bayesian inference. The surrogate model is constructed via a regularized alternative least-square regression with Tikhonov regularization using a roughening matrix computing the gradient of the solution, in conjunction with a perturbation-based error indicator to detect optimal model complexities. The  model approximates a vector of a continuous solution at discrete values of a physical variable. The required number of random realizations to achieve a successful approximation linearly depends on the function dimensionality. The computational cost of the model construction is quadratic in the number of random inputs, which potentially tackles the curse of dimensionality in high-dimensional stochastic functions. Furthermore, this vector valued separated representation-based model, in comparison to the available scalar-valued case, leads to a significant reduction in the cost of approximation by an order of magnitude equal to the vector size. The performance of the method is studied through its application to three numerical examples including a 41-dimensional elliptic PDE and a 21-dimensional cavity flow. 
\end{abstract}


\end{frontmatter}


%
\section{Introduction}
\label{sec: Introduction}

An inverse problem arises when the inputs/complexities of a model are estimated indirectly from outputs, e.g., noisy observations \cite{Vogel02,Ma09b,Tarantola05,Constantine11a,Beck09,Daubechies04,Tropp10a}. In this context, the Bayesian approaches, which have recently attracted much attention \cite{Bernardo94,Gelman03,Spall10}, provide applied probability and uncertainty measurements for statistical inference. Indeed, as an extention of conventional statistical methods \cite{Kaipio04,Charles93,Hansen10,Vogel02}, the solution of the Bayesian inference is a posterior probability distribution over the model input/complexities regarding available/unavailable prior knowledge about them  \cite{Aster05}. The computational cost of estimating the posterior distribution is a challenge in practice, and, in response, many asymptotic, deterministic, and sampling based methods have been developed focusing on reductions of or surrogates the forward model \cite{Tarek12,Marzouk09b,Marzouk09a,Youssef09,Jaakkola2000}. 

Deterministic methods might be reasonable alternatives in low to moderate dimensions, but for high-dimensional and complex problems, the Markov Chain Monte Carlo (MCMC) \cite{Metropolis53} strategy is a more general and flexible approach \cite{Bremaud99,Tierney94,Gilks96,Dave}. The MCMC approach requires evaluation of the likelihood function \cite{Miller04}, indeed, solving the forward model many times, which might be costly and/or intractable. In the case of intensive computational models, e.g., those described by a system of Ordinary Differential Equations (ODEs) or Partial Differential Equations (PDEs), the cost of such an approach becomes prohibitive. To do so, generalized polynomial chaos (gPC)-based \cite{Wan05}, Stochastic Galerkin, and Collocation methods \cite{Marzouk09b,Doostan07} have been developed. However, these approaches are attractive alternatives for low (or moderate) dimensions. In the case of high-dimensional problems, \textit{low-rank non-intrusive separated representation approximation} of the model is proposed to construct a surrogate of the forward model \cite{Beylkin02,Beylkin05,Ammar06,Doostan12}. Then, for efficient Bayesian inference, this surrogate model is used in computing the likelihood function, as well as the posterior probability distribution function.

In 1927, Hitchcock \citep{Hitchcock} introduced separated representation, which is also known as parallel factor analysis or canonical decomposition, to express a Polyadic as a sum of products of rank-one vectors. Subsequently, this approach has been widely used in a variety of areas including chemical kinetics \citep{Doostan12}, data mining \citep{Kolda09a,Kolda09b,Hackbusch04,Kroonenberg80}, and image processing \citep{Furukawa02,Shashua01}. Here, an approach is proposed to construct a vector valued separated representation of a continuous stochastic function of a physical random variable $\xi$ and input random variables $\bm{y}$, i.e., $\bm{u}\left(\xi,\bm{y}\right)$, $\bm{y}=\left(y_1,\ldots, y_d\right)$, $d\in\mathbb{N}$. This function can be approximated with accuracy $\mathcal{O}\left(\bm{\epsilon}\right)$ in a separated form as
\begin{equation}
\label{eq: separated representation in introduction}
\bm{u}\left(\xi,\bm{y}\left( w \right) \right)=\sum_{l=1}^{r} s_l \bm{u}_0^l(\xi) \prod_{i=1}^du_i^l(y_i(w))+\mathcal{O}\left(\bm{\epsilon}\right),
\end{equation}
where $\bm{u}_0^l(\xi)$, which is a vector valued univariate function of a physical variable $\xi$; $\left\lbrace u_i^l(y_i(w)) \right\rbrace_{i=1}^d$, $l=1,\ldots,r$, which are univariate functions of random variables; and $s_l$, which are normalization constants; are unknown and must be computed. Because the separation rank, $r$, as one of the model complexities, is independent of problem dimensionality \citep{Beylkin08,Beylkin05,Beylkin02,Doostan12}, $d$, the computational complexity is a weakly-linear function in $d$, which remarkably reduces the curse of dimensionality, a bottleneck for uncertainty quantification of high-dimensional functions. Furthermore, the model has low-rank separated representation approximation structure if a small separation rank can be found for it.

This study is organized as follows. In section \ref{sec:Bayesian Inference}, the Bayesian inference is discussed in more detail. In section $\ref{sec:Problem Setup}$, the general problem setup described for either system of ODEs or PDEs. Thereafter, in section \ref{sec: Separated Representation}, the vector case of the separated representation is introduced, and in sections \ref{sec:Regularization} and \ref{subsec: Perturbation based error indicator}, a regularization approach and an error indicator are proposed for stabilizing the method and finding the optimum construction of the separated model. In section $\ref{sec: Results}$, the results are presented for three different examples: a manufactured function, an elliptic equation, and a cavity flow problem.

\section{Bayesian Inference}
\label{sec:Bayesian Inference}

The goal of an inverse problem is to recover anterior information from available data \cite{Constantine11a,Vogel02,Beck09}. The quantity of interest, $\bm{u}$, in the forward problem context is computed given a mathematical model, $\mathcal{A}$, and parameters, $\bm{y}$; however, in the inverse problem either the parameters or the mathematical model is computed given the other two quantities.

Considering a general system of equations $\mathcal{A}(\bm{y}) \approx\bm{u}$, there are two main approaches for parameter estimation: main classical least squares\footnote{The classical least square approaches are linear and non-linear regression, and data free inference \cite{Vogel02,Andreas96}.} and Bayesian strategies. In the Bayesian approaches the model is treated as a random variable and the solutions are  probability distributions for those model parameters that are sought \cite{Aster05}. Detailed statistical knowledge (e.g., mode, mean, standard deviation, correlation, smoothness, etc.) of the parameters can be revealed using the probability distributions, while in the classical methods the solutions are point quantities and the parameter statistics are not available. In Bayesian approaches, the prior information, which comes from other sources (e.g., physical and experimental observations), is called the $``prior \ distribution"$ of parameters $\bm{y}$, and is denoted by $p\left(\bm{y}\right)$. The $``posterior \ distribution"$, $q\left(\bm{y}|\bm{u}\right)$, can be formulated by incorporating the given data along with the prior distribution in Bayes' theorem \cite{Bayes1763,stephen06} as follows:
\begin{equation}
\label{eq:Posterior distribution}
q\left(\bm{y}|\bm{u}\right)=\frac{p\left(\bm{u}|\bm{y}\right)p\left(\bm{y}\right)}{\int p\left(\bm{u}|\bm{y}\right)p\left(\bm{y}\right)d\bm{y}}.
\end{equation}

The data are incorporated in the formulation through the likelihood function $p\left(\bm{u} \right)$, which can be presented as $L(\bm{y}) \equiv p\left(\bm{u}|\bm{y}\right)$.

\begin{rem}
\label{rem:prior and post dist}
In reality, prior and posterior distributions show the strength of perception about feasible values for the inputs, $\bm{y}$, before and after experiencing the outputs, $\bm{u}$. More prior information on the model parameters, e.g., a range of possible values, leads to a more suitable prior distribution. If there is no available prior information about the parameters, then based on ``the principle of indifference'' an  ``uninformative'' prior distribution is chosen; in which all the model parameter values are assumed to have the same likelihood.
\end{rem}

In general, determining the posterior distribution is computationally expensive and problematic due to the integral in (\ref{eq:Posterior distribution}), which is usually a high-dimensional integral. A typical simplified model is assumed when the  value of the integral is not really needed. In these situations two different model posterior distributions are compared by computing the likelihood functions; therefore  Eq. (\ref{eq:Posterior distribution}) can be written as
\begin{equation}
\label{eq:simplified Posterior distribution}
q\left(\bm{y}|\bm{u}\right) \propto p \left(\bm{u}|\bm{y}\right)p\left(\bm{y}\right).
\end{equation}

The general system of equations is converted to 
\begin{equation}
\label{eq:system of eq}
\bm{u}=\mathcal{A}(\bm{y})+\bm{\eta},
\end{equation} 
where $\bm{\eta}$ is assumed to be an independent and identically distributed (i.i.d) noise vector of size $n$, which is normally distributed with zero mean, $\bm{\sigma}$ standard deviation, and $p_{\eta}$ noise density, i.e., $\eta_k\sim N\left(0,\sigma_k^2\right),\ k=1,\ldots,n$. Here, $\bm{\eta}$ may cover both experimental and modelling errors. The modelling error encompasses numerical errors and the errors due to simplifying assumptions neglecting some physics of the problem. The likelihood function $L\left(\bm{y}\right)$ can be presented as:
\begin{equation}
\label{eq:liklihood function}
L\left(\bm{y}\right)\equiv \prod_{j=1}^{N} p_{\eta} \left( \bm{u}^{(j)}- \mathcal{A}^{(j)}\left(\bm{y}\right) \right).
\end{equation}
Therefore, the posterior distribution can be written as 
\begin{equation}
\label{eq:final posterior distribution}
q\left(\bm{y}|\bm{u}\right) \propto L\left(\bm{y}\right)p\left(\bm{y}\right),
\end{equation}
by comparing equations (\ref{eq:liklihood function}) and (\ref{eq:simplified Posterior distribution}).

\begin{rem}
\label{rem:normally distributed forward model}
In the case of i.i.d noise in the measured data,  $\mathcal{A}\left(\bm{u}|\bm{y}\right)$ is normally distributed with $\bm{u}$ mean and $\bm{\sigma}$ standard deviation, i.e., $\mathcal{A}\left(\bm{u}|\bm{y}\right)\sim N\left(\bm{u},\bm{\sigma}^2\right)$.
\end{rem}

The characteristics of the posterior distributions can be computed by several methods including numerical integration, asymptotic approximation, and sampling-based approaches \cite{Haario04,McKeague05}. The sampling-based approaches generate samples many times to examine whether the prior distribution approximates the posterior distribution. Several sampling methods are available to explore posterior distributions \cite{Geweke89,Geweke92,Stewart79,Zellner84,Smith84} for low and moderate-dimensional problems. Because this work deals with high-dimensional problems, the Markov Chain Monte Carlo strategy (MCMC) of Bayesian inference is used along with the combination of two powerful ideas: Delaying Rejection (DR) \cite{Tierney94,Mira02,GreenP01} and Adaptive Metropolis (AM) sampling \cite{Haario99b,Heikki98}, which together are known as DRAM \cite{Heikki06}.

In the DRAM approach, the forward model may need to be computed, e.g., $10^6$ times, which might be very expensive. In order to decrease the computational cost of the simulations, various methods such as generalized polynomial chaos, Stochastic Galerkin, and Collocation were developed to approximate the output \cite{Wan05,Marzouk09b,Doostan07}. The cost function of those proposed models is exponentially proportional to the problem dimension; therefore they are adequate for low to moderate-dimensional problems. Here, for high-dimensional problem cases, a surrogate model in the context of  separated representation \cite{Beylkin02} is developed, in which the computational cost is a quadratic function of the dimensionality. In section \ref{sec: Separated Representation}, the separated representation approximation is explained in detail. In the next section, the problem setup is introduced for the PDE/ODE system of equations.

\section{Problem Setup}
\label{sec:Problem Setup}

Let $\left( \Omega,\mathcal{F},\mathcal{P} \right)$ be a complete probability space, where $\mathcal{F}$ is the $\sigma$-algebra of events, $\Omega$ is the set of elementary events, and  $\mathcal{P}:\mathcal{F}\rightarrow [0,1]$ is a probability measure on $\sigma-$field $\mathcal{F}$. A generic stochastic Partial/Ordinary Differential Equation (PDE/ODE) can be formulated as
\begin{equation} \label{eq:system of ODE}
\mathcal{A} \left(\xi,\bm{y}(\omega)  ;\bm{u} \right)=0,\ \ \left(\xi, \omega\right) \in \left[\Xi_1,\Xi_2\right] \times\Omega  ,
\end{equation}
where $\mathcal{A}$ defines the forward model. $\xi \in \left[\Xi_1,\Xi_2\right], \ \left(\Xi_1,\Xi_2\right) \in \mathbb{R}\times \mathbb{R}$, is a physical (spatial/temporal) variable and $\bm{y}(\omega)=\left(y_1(\omega),\ldots, y_d(\omega)\right):\Omega\rightarrow \mathbb{R}^{d}$, $d\in\mathbb{N}$, is a vector of random inputs contaminated by uncertainties. According to a probability density function of $y_i$,  $\rho(y_i):\Gamma \subseteq \mathbb{R} \rightarrow \mathbb{R}_{\geq 0}$, where $i$ varies from $1$ to $d$, the components of the random vector $\bm{y}(\omega)$ are assumed i.i.d. $\bm{u}$ is the continuous solution, but, here, a discrete approximation or a vector valued solution at $n$ different values of $\xi$ is considered, i.e., $\bm{u}: \mathbb{R}^{d+1} \rightarrow \mathbb{R}^{n}$. Indeed, the solution of interest can be shown as: 
\begin{equation}
\label{eq:solution}
\bm{u}\left(\xi,\bm{y}\left( w \right) \right):= \bm{u}\left(\xi, y_1\left( w \right), \ldots , y_d \left( w \right)\right) : \left[\Xi_1,\Xi_2 \right] \times \Gamma^d\rightarrow \mathbb{R}^{n}.
\end{equation}
Furthermore, an appropriate boundary conditions and initial values are considered related to the problems introduced by (\ref{eq:system of ODE}).
\section{Separated Representation}
\label{sec: Separated Representation}
Separated representation techniques in high-dimensional function approximations, potentially eliminate the curse of dimensionality by approximating a $d$-dimensional function by solving $d$ one-dimensional functions \cite{Beylkin02,Doostan12,Beylkin05}. In this section the separated representation heuristic, an algorithm for the technique, and core principles are reviewed. To avoid instability, a Tikhonov regularization is introduced and to detect the optimal model structure, a perturbation-based error indicator is defined.

The goal is to non-linearly estimate the vector valued functions $\bm{u}\left(\xi,\bm{y}\left( w \right) \right)$ using the equivalent separated representation as:
\begin{equation}
\label{eq: separated representation}
\bm{u}\left(\xi,\bm{y}\left( w \right) \right)=\sum_{l=1}^{r} s_l \bm{u}_0^l(\xi) \prod_{i=1}^du_i^l(y_i(w))+\bm{\varepsilon}.
\end{equation}

Here, $r \in \mathbb{N}$, the separation rank, is not given \textit{a priori} and is estimated by the defined error indicator, which is described in section \ref{subsec: Perturbation based error indicator}. $\bm{u}_0$ is a vector valued univariate function of a temporal/spatial variable and $\left\lbrace u_i^l(y_i) \right\rbrace_{i=1}^d \in \mathbb{R}, \ l=1,\ldots,r$ are univariate functions of random input variables $y_i$. $\lbrace s_l \rbrace_{l=1}^r \in \mathbb{R}_{>0}$ are scalar normalization values. Similar to the separation rank, these values are not known \textit{a priori} and must be computed. 

\begin{define}
The space of $r$-separation rank and $d+1$-dimensional functions is 
\begin{equation}
\label{eq:separated space}
\bm{\mathcal{U}}_r=\left\lbrace  \sum_{l=1}^{r} s_l \bm{u}_0^l \prod_{i=1}^du_i^l(y_i) \right\rbrace.
\end{equation}
\end{define}
\begin{define}
Given a set of $N$ independent random inputs $\bm{y}^{(j)},  j=1,\ldots,N$, and the corresponding vector valued solutions  with size $n$, the data set $D$ is defined by
\begin{equation}
\label{eq:Data realization}
D=\left\lbrace \left( \bm{y}^{(j)}; \bm{u}^{(j)}\left(\xi,\bm{y}^{(j)} \right)\right)\right\rbrace_{j=1}^N,
\end{equation}
and the Frobenius norm of $\bm{u}$ is formulated as 
\begin{equation}
\label{eq:Frobenius norm}
\Vert \bm{u} \Vert_D = \left\langle \bm{u}\ ,\bm{u}  \right\rangle_D^{\frac{1}{2}},
\end{equation}
where the inner product between $\bm{u}$ and $\bm{v}$ can be defined such that
\begin{eqnarray}
\label{eq:inner product}
\left\langle \bm{u}\ ,\bm{v}  \right\rangle_D &=& \left\langle \left\lbrace \bm{y}^{(j)} \ , \bm{u}^{(j)} \left(\xi,\bm{y}^{(j)} \right) \right\rbrace_{j=1}^N \ , \left\lbrace \bm{y}^{(j)} \ , \bm{v}^{(j)} \left(\xi,\bm{y}^{(j)} \right) \right\rbrace_{j=1}^N \right\rangle_D \\
&=& \frac{1}{n N} \sum_{j=1}^N \bm{u}^{(j)} \cdot \bm{v}^{(j)}.
\nonumber
\end{eqnarray}
\end{define}

The separated approximation of $\bm{u}$, $\bm{u}_s$, can be estimated via the solution of a least-squares regression problem
\begin{equation}
\label{eq:least squares regression}
\bm{u}_s=\arg\min_{\hat{\bm{u}}_s\in\bm{\mathcal{U}}_r}\ \Vert \bm{u} - \hat{\bm{u}}_s\Vert_D^2 .
\end{equation}

The current non-linear schemes to solve non-linear optimization problems (\ref{eq:least squares regression}), e.g., damped Gauss-Newton \cite{Bjorck96}, are prohibitively expensive for high-dimensional problems and are limited to low (or moderate)-dimensional problems. Alternatively, for high-dimensional problems, the multi-linear alternating least squares (ALS) method \cite{Rao99} is used. In this  approach, at the separation rank $l$, the univariate function along dimension $k$, ${u_k^l(y_k)}$, is solved by constructing the related one-dimensional least-squares regression problem, and freezing the other univariate functions ${u_i^l(y_i)}, i=1,\ldots,d, \ i \neq k$, at their current values. The regression process is repeated for each dimension in turn till all the univariate functions are solved.

With respect to the probability density functions of $y_i$, $\rho(y_i)$, the univariate functions $u_i^l(y_i)$ are expanded into a finite dimensional bases, e.g., orthogonal spectral polynomials in the case of Polynomial Chaos Expansion (PCEs). These functions are approximated by
\begin{equation}
\label{eq: univariate function expansion}
u_i^l\approx \sum_{\alpha=0}^M c_{\alpha,i}^l \psi_{\alpha} \left(y_i\right),
\end{equation}
where $\left\lbrace \psi_\alpha \left( y_i \right) \right\rbrace$ is a set of spectral (e.g., Legendre and Hermit) polynomials of degree $\alpha \leq M \in \mathbb{N}_0 := \mathbb{N}\cup \left\lbrace 0 \right\rbrace$. The expansion coefficients $\bm{c}_i:= \left(c_{0,i}^1,\ldots,c_{M,i}^1,\ldots,c_{0,i}^r,\ldots,c_{M,i}^r \right) \in \mathbb{R}^{r(M+1)}$ along dimension $i=1,\ldots,d$, are computed by reducing (\ref{eq:least squares regression}) to a discrete least-squares optimization such as
\begin{eqnarray}
\label{eq: equivalent eq of least square eq}
\ {\bm{c}}_i 
&=&\arg\min_{\hat{\bm{c}}_i}\ \Vert \bm{u} - \hat{\bm{u}}_r\Vert_{D}^2 \\
&=& \arg\min_{\hat{\bm{c}}_i}
 \left\Vert \bm{u} -\bm{u}_0 \sum_{l=1}^r \left( \sum_{\alpha =0}^M \hat{c}_{\alpha,i} ^l \psi_{\alpha}\left( y_i\right)  \right) s_l \prod_{k=1\neq i}^d u_k^l \left( y_k\right) \right\Vert_D^2.
\nonumber
\end{eqnarray} 

Here $\hat{\bm{u}}_r$ is a ranked $r$ approximation of $\bm{u}$. 

The first derivative of (\ref{eq:least squares regression}) with respect to the random input variables is set to zero to compute the expansion coefficients, which leads to solve the following system of system of equations:
\begin{equation}
\label{eq:normal equation}
\bm{A}^T_i \bm{A}_i\bm{c}_i=\bm{A}^T_i\bm{u}, \  i=1,\ldots,d.
\end{equation}
The matrix $\bm{A}_i \in \mathbb{R}^{(n  N)\times (r  \left( M+1 \right) )}$ is a column block structured matrix, $\bm{A}_i = \left[ \bm{A}_i^1 \ldots \bm{A}_i^r\right]$. Each column-block matrix $\bm{A}_i^l \in \mathbb{R}^{(n  N)\times \left( M+1 \right)}$, is computed by 
\begin{eqnarray}
\label{eq:column block matrix for i=1:d}
\bm{A}_i^l\left((j-1) n+1 :(j-1) n+n,\alpha+1 \right)\\
=\bm{u}_0 s_l \psi_{\alpha} ( y_i^{\left( j\right)})\prod_{k=1\neq i}^d u_k^l ( y_k^{\left( j\right)}), \ i=1,\ldots,d.\nonumber
\end{eqnarray}

Because the function $\bm{u}$ may not be a smooth function of the physical random variable, $\bm{u}_0$ is directly solved without expanding it into a spectral polynomial. The equivalent relations of (\ref{eq: equivalent eq of least square eq}), (\ref{eq:normal equation}), and (\ref{eq:column block matrix for i=1:d}) for solving $\bm{u}_0$ are 
\begin{eqnarray}
\label{eq: equivalent eq of least square eq for u0}
\ {\bm{u}}_0 
&=& \arg\min_{\hat{\bm{u}}_0}\ \Vert \bm{u} - \hat{\bm{u}}_r\Vert_{D}^2\\
&=&\arg\min_{\hat{\bm{u}}_0} \left\Vert \bm{u} - \bm{u}_0 \sum_{l=1}^r s_l \prod_{i=1}^d u_i^l \left( y_i\right) \right\Vert_D^2,\nonumber 
\end{eqnarray}
\begin{equation}
\label{eq:normal equation for u0}
\bm{A}_0\bm{u}_0=\bm{u},
\end{equation}
and
\begin{eqnarray}
\label{eq:column block matrix for u0}
\ \bm{A}_0^l\left((j-1) n+1 :(j-1) n+n,\alpha+1 \right)\\
=\bm{u}_0 s_l \prod_{i=1}^d u_i^l ( y_i^{\left( j\right)}),\ \ \bm{A}_0^l \in \mathbb{R}^{(n  N)\times (r \left( M+1 \right) )},\nonumber
\end{eqnarray}
respectively.

The non-intrusive separated representation approximation, similar to other regression methods, may suffer from the issue of instability for the given complexity parameters; therefore, Tikhonov regularization is utilized here.

\subsection{Regularization}
\label{sec:Regularization}

In each iteration of the ALS algorithm that the unknowns of the separated representation formulation are updated, the residual norm $\Vert\bm{u}-\bm{u}_s\Vert_D$ decreases. Model structures with larger values of ($r,M$) lead to a greater decrease of the residual norm; therefore, one may expect that the larger the values of the model complexities, the more accurate the results. However, in the cases of non-separable functions, lack of information, or noisy data, excessive reduction of the residual norm results in instability, in which the method can match the realization solutions individually but be completely unreasonable for other data points. A naive parametric approach to avoid this issue is choosing small values for $(r,M)$, which may lead to unfitted approximation. Instead, a non-parametric approach is used based on the concept of regularization by encouraging additional smoothness constraints on the approximated solution, $\bm{u}_s$. For a given $r$ and $M$ a Tikhonov regularization \cite{Aster05} is examined by adding a smoothness penalty term $\Vert \bm{L} \bm{c} \Vert_2^2$, $\bm{L} \in \mathbb{R}^{(r  \left( M+1 \right)) \times (r  \left(M+1 \right))}$, to the regression cost function (\ref{eq: equivalent eq of least square eq}), i.e.,
\begin{equation}
\label{eq: regularized regression c}
\ {\bm{c}}_{reg} = \arg\min_{\hat{\bm{c}}_{reg}}\ \frac{1}{n N}\ \Vert \bm{A}\hat{\bm{c}}_{reg} - \bm{u}\Vert_2^2 + \lambda^2 \Vert \bm{L} \hat{\bm{c}}_{reg} \Vert_2^2 ,
\end{equation}
where $\bm{L}$ is a roughening matrix, $\lambda \in \mathbb{R}_{\geq 0}$ is a regularization parameter, and $\bm{c}_{reg}$ is a matrix of the expansion coefficients.

Selecting suitable values for $\lambda$ and $\bm{L}$ is essential for better performance of the Tikhonov regularization. Among several available statistical methods to estimate the $\lambda$, e.g., Morozov's Discrepancy Principle, L-curve, Predictive Risk Estimator, and  Generalized Cross Validation (GCV) \cite{Morozov93,Hansen10,Aster05}, the \textit{GCV} is found to be more accurate and used in this study. The value of $\lambda$ in comparison to the singular values of $\bm{A}$ is important to the question of whether to regularize the problem. Consider $\varsigma_j$, where $j=1,\ldots,nN$, as a singular values of Singular Value Decomposition (SVD) of matrix $\bm{A}$, such that $\varsigma_1 \geq \varsigma_2 \geq \ldots \geq \varsigma_{nN}$. Intuitively, if the value of the regularization parameter, $\lambda$, is close to the minimum singular value $\varsigma_{nN}$, i.e.,
\begin{equation}
\label{eq:decide regularization}
\lambda \leq C \times \varsigma_{nN},\  C \in \mathbb{R}_{\left[0.99\  1.01\right]},
\end{equation}
the problem is not ill-posed and the penalty term just adds noise to it; therefore, the regularization is not needed. In each iteration of the ALS process this condition is checked to decide whether to regularize the problem.

The other important factor of the Tikhonov regularization is  the roughening matrix, which  affects the effectiveness and performance of the method. In the standard form of the regularization, where the roughening matrix is an identity matrix ($\bm{L}={I}$), the objecting function involving $\Vert \bm{c} \Vert_2$ (and $\Vert \bm{u}_0 \Vert_2$) is minimized. In highly sparse regions, where large deviations are penalized in the reconstruction, the standard Tikhonov regularization favors inaccurate solutions. Therefore, a roughening matrix based on the solution gradient is derived as follows
\begin{equation}
\label{eq:roughening matrix c}
\Vert \bm{L} \bm{c} \Vert_2^2 = \mathbb{E} \left[ \nabla \bm{u}_s^T \nabla \bm{u}_s \right],
\end{equation}
where $\nabla \bm{u}_s$ is the gradient of the separated approximation of $\bm{u}$ with respect to the $y_i$, $\ i=1,\ldots,d$. In comparison to the work of \cite{Doostan12}, where the second order moment of the solution is considered, $\Vert \bm{L} \bm{c} \Vert_2^2 = \mathbb{E} \left[ \bm{u}_s^2 \right]$, the value of (\ref{eq:roughening matrix c}) is much larger, which can promote smoothness and more heavily penalize coefficients corresponding to higher order polynomials. In practice, as will be demonstrated by the numerical examples in section \ref{sec: Results}, less instability and more control over the solution can be expected.

It is straightforward to show that $\mathbb{E} \left[ \nabla \bm{u}_s^T \nabla \bm{u}_s \right]=\bm{c}^T \bm{B} \bm{c}$, where $\bm{B}=\bm{L}^T\bm{L} \in \mathbb{R}^{(r(M+1))\times (r(M+1))}$ is a positive definite and symmetric matrix. Each $\left(l,l' \right)$-th block of $\bm{B}$ is computed by the following equation:
\begin{equation}
\label{eq:betta for roughening matrix c}
\bm{B} \left(l,l'\right)=s_ls_{l'}\left(\bm{u}_{0_l}^T  \bm{u}_{0_l} \right) \prod_{i\neq k}^{d} \left( \sum_{\alpha =0}^M c_{\alpha ,i}^l c_{\alpha ,i}^{l'} \right) \left( \sum_{\alpha =0}^M \sum_{\alpha '=0}^M c_{\alpha ,k}^l c_{\alpha ,k}^{l'} \gamma_{\alpha \alpha '} \right) \bm{I}_{M+1},
\end{equation}
where $\gamma_{\alpha \alpha '} = \left\langle \nabla \bm{\psi}_{\alpha} \nabla \bm{\psi}_{\alpha '} \right\rangle$ and $\bm{I}_{M+1}$ denotes an identity matrix of size $M+1$. The gradient of discretized $\bm{u}_0$ can be approximated by $\bm{L}_0\bm{u}_0$, where
\begin{equation}
\bm{L}_0=\left[
\begin{tabular}{c c c c c}
 -1 &  1 &        &    &    \\ 
    & -1 & 1      &    &    \\ 
    &    & \ldots &    &    \\ 
    &    & -1     & 1  &    \\ 
    &    &        & -1 & 1  \\ 
    
\end{tabular}
\right].
\end{equation}

Similarly, the equivalent equation of (\ref{eq: regularized regression c}) for $\bm{u}_0$ is given by
\begin{equation}
\label{eq:eq: regularized regression u0}
\ {\bm{u}}_{0_{reg}} = \arg\min_{\hat{\bm{u}}_{0_{reg}}}\ \frac{1}{nN}\ \Vert \bm{A}_0\hat{\bm{u}}_{0_{reg}} - \bm{u}\Vert_2^2 + \lambda^2 \Vert \bm{L}_0 \hat{\bm{u}}_{0_{reg}} \Vert_2^2.
\end{equation}
%

The issue of instability could be mitigated by regularization, but still there is a need to detect the optimal $r$ and $M$ to tackle the issue of over/under-fitting, when the complexity parameters are greater or smaller than the optimal values. In the next section, an indicator based on a perturbation error is introduced to detect the optimal model complexities.

\subsection{Perturbation based error indicator}
\label{subsec: Perturbation based error indicator}

To detect the optimal values of separated rank and polynomial degree, $\left(r,M \right)$, an error indicator is introduced by defining a perturbation bound on the sensitivity of both regularized and non-regularized solutions (Equations: \ref{eq: equivalent eq of least square eq}, \ref{eq: equivalent eq of least square eq for u0}, \ref{eq: regularized regression c}, and \ref{eq:eq: regularized regression u0}). For regularized cases and $i=1,\ldots,d$, the Perturbation-based Error Indicator (PEI) can be derived as
\begin{equation}
\label{eq:error indicator}
PEI_i= \frac{\left(n N\right)^{0.5} \lambda_i^{-1} \Vert \bm{L}^{-1} \Vert_2 \ \hat{\sigma}_i}{\Vert \bm{c}_i \Vert_2},\ \ i=1,
\ldots,d ,
\end{equation}
where $\bm{c}_i$ is the solution of the problem (\ref{eq: equivalent eq of least square eq}) and $\lambda_i$ is the regularization parameter for dimension $i$ \cite{Vogel02}. For non-regularized cases, where $\lambda_i=0$, the minimum singular value of matrix $\bm{A}$ in (\ref{eq:error indicator}) is used instead. It is assumed that the separated representation errors, $\bm{\varepsilon}$, in (\ref{eq: separated representation}) can be approximated by random variables with zero means and $\bm{\sigma}$ standard deviations. Here, $\hat{\sigma}_i$ is the estimation of the standard deviation $\sigma_i$, which is given by
\begin{equation}
\label{eq:sigma hat in error indicator}
\hat{\sigma}_i = \frac{\Vert \bm{A} \bm{c}_i-\bm{u} \Vert_2^2}{n N-tr\left(\bm{H}_i \right) }.
\end{equation}
The \textit{hat matrix}, $\bm{H}_i=\bm{A}\left( \bm{A}^T\bm{A}+\lambda_i^2 \bm{L}^T \bm{L} \right)^{-1} \bm{A}^T$, is a mapping matrix of the realizations to their separated representation approximations. 

For the $0$-th dimension the equivalent error indicator is derived as
\begin{equation}
\label{eq:error indicator for u0}
PEI_0= \frac{\left(n N\right)^{0.5} \lambda_0^{-1} \Vert \bm{L}_0^{-1} \Vert_2 \ \hat{\sigma}_0}{\Vert \bm{u}_0 \Vert_2},
\end{equation}
where $\lambda_0$ is a regularization parameter for dimension $d+1$.

 For each pair of $\left(r,M \right)$ there is a $(d+1)$ size PEI vector associated with the last iteration of the ALS. For that particular pair, the maximum value of the PEI vector, $PEI_{max}^{(r,M)}$, is chosen. Notice that the PEI depends on  $r$ and $M$ indirectly through $\bm{c}$, and conservatively $\lambda$ and $\bm{L}$. The PEI associated with unnecessarily small/large model complexities is a large value. However, it is relatively small for those model complexities which could be optimal, e.g., where the standard deviation error is minimal. Among all possible model structures, the one which corresponds to the minimum value of $PEI_{max}^{(r,M)}$ is selected as the optimal separated representation model. In algorithm (\ref{Algorithm:ALS}), the overall non-intrusive ALS procedure including the proposed regularization strategy and perturbation error indicator is summarized.

\subsection{Computational cost}
\label{Computational cost}

It is worthwhile to elaborate on the computational cost and complexity of the algorithm. For a given $y_i$, each univariate function $u_i^l(y_i)$, is evaluated with  complexity $\mathcal{O}(MK)$, where $K$ is the number of ALS iterations. Therefore, the cost of computing matrix $\bm{A}$ in (\ref{eq:normal equation}) is $\mathcal{O}\left( rMdKNn\right)$. For a full sweep of the ALS, all the normal equations (\ref{eq:normal equation}) can be solved using Cholesky decomposition of $\bm{A}^T \bm{A}$ with complexity $\mathcal{O}\left( r^2M^2dK^2Nn^2\right)$ for each. Similarly, equation (\ref{eq:error indicator for u0}) can be computed with cost $\mathcal{O}(r^2K^2Nn)$. 

It is assumed that $N \gg rM(d+1)$, which is an asymptotic but relevant assumption. Therefore, the complexity of the algorithm is $\mathcal{O}\left( r^3M^3d^2K^2n^2\right)$, which is quadratic in $d$. For the situations where the forward model is expensive or the cost of a surrogate model exponentially grows as a function of dimensionality, the separated approximation with this cost is an outstanding success. In the next section, the results of numerical examples are provided.

\begin{algorithm}[H]
\SetKwData{Left}{left}
\SetKwData{This}{this}
\SetKwData{Up}{up}
\SetKwFunction{Union}{Union}
\SetKwFunction{FindCompress}{FindCompress}
\SetKwInOut{Input}{ Input}
\SetKwInOut{Output}{ Output}
$\bullet$\Input{Data set $D=\left\lbrace\left( \bm{y}^{(j)}; \bm{u}(\bm{y}^{(j)})\right)\right\rbrace_{j=1}^N$ and accuracy $\epsilon$}
\BlankLine
$\bullet$\Output{$r$, $M$, $\{c_{\alpha ,i}^l\}_{i=1}^d$, and $s_l$ for $\alpha=0,\dots,M$ and $l=1,\dots,r$}
\BlankLine
$\bullet$ Set $r = 1$ and $M=1$ and initialize $\bm{c}_i$ and $\bm{u}_0$ randomly
\BlankLine
	\While{$\Vert \bm{u}-\bm{u}_s\Vert_D > \epsilon$}{
	\BlankLine
		\While{$\Vert \bm{u}-\bm{u}_s\Vert_D$ decreases much}{
			\BlankLine
			\For{$\alpha \leftarrow 0$ \KwTo $M$}{
			\BlankLine
			$\bullet$ Fix $\lbrace c_{\alpha,i}^l \rbrace_{i=1}^d$ and solve $\bm{u}_0$ using: 
			$\left\lbrace \begin{tabular}{c c}
			$Eq. (\ref{eq: equivalent eq of least square eq for u0} )$ & if $(\ref{eq:decide regularization}) \ is \ \textit{.True.}$\\
			$Eq. (\ref{eq:eq: regularized regression u0})$ & if $(\ref{eq:decide regularization}) \ is \ \textit{.False.} $			
			\end{tabular} \right\rbrace$									             \BlankLine
			$\bullet$ Update $s_l\leftarrow s_l\  \Vert 	\bm{u}_0\Vert_D$ and $\bm{u}_0 \leftarrow 	\bm{u}_0/ \Vert \bm{u}_0 \Vert_D$,
			\BlankLine
	 		\For{$i\leftarrow 1$ \KwTo $d$}{
			\BlankLine
			$\bullet$ Fix $\lbrace c_{\alpha,k}^l \rbrace_{k\ne i}$ and solve $\lbrace c_{\alpha,i}^l\rbrace$ using:	 $\left\lbrace \begin{tabular}{c c}
			$Eq. (\ref{eq:least squares regression} )$ & if $(\ref{eq:decide regularization}) \ is \ \textit{.True.}$\\
			$Eq.(\ref{eq: regularized regression c})$ & if $(\ref{eq:decide regularization}) \ is \ \textit{.False.} $
			\end{tabular} \right\rbrace$	\\
			\BlankLine
			$\bullet$ Update $s_l\leftarrow s_l\  \Vert 	\bm{u}_i^l\Vert_D$ and $c_{\alpha,i}^{l}\leftarrow 	c_{\alpha,i}^{l}/ \Vert \bm{u}_i^l\Vert_D$
			\BlankLine
			}	
			}	
  	     \BlankLine
		}
		\BlankLine
		$\bullet$ Set $r = r+1$ and (randomly) initialize $\bm{c}_i^r$ for $i=1,\dots,d$, and $\bm{u}_0^r$
		\BlankLine			
	$\bullet$ If $rMd > N$ set $r=1$ and $M = M+1$; (randomly) initialize $\bm{c}_i$ and $\bm{u}_0$ 
}
$\bullet$ Report the optimum $r$ and $M$ based on the error indicator and correspondence $\bm{c}$, $\bm{u}_0$, and $s_l$
\label{Algorithm:ALS}
\caption{The algorithm for constructing a separated representation approximation model of a vector valued stochastic function, non-intrusively.}
\end{algorithm}
\section{Results}
\label{sec: Results}
In this section, the performance of the non-intrusive separated representation model and its application in Bayesian inference is investigated through a $11$-dimensional manufactured function, a 1D in space $41$-dimensional elliptic PDE, and a 2D in space $21$-dimensional cavity flow. 

\subsection{A manufactured function}
\begin{figure}
    \centering
    \begin{tabular}{cc}
            \hspace{-0.5cm}    
      \includegraphics[width=2.8in]{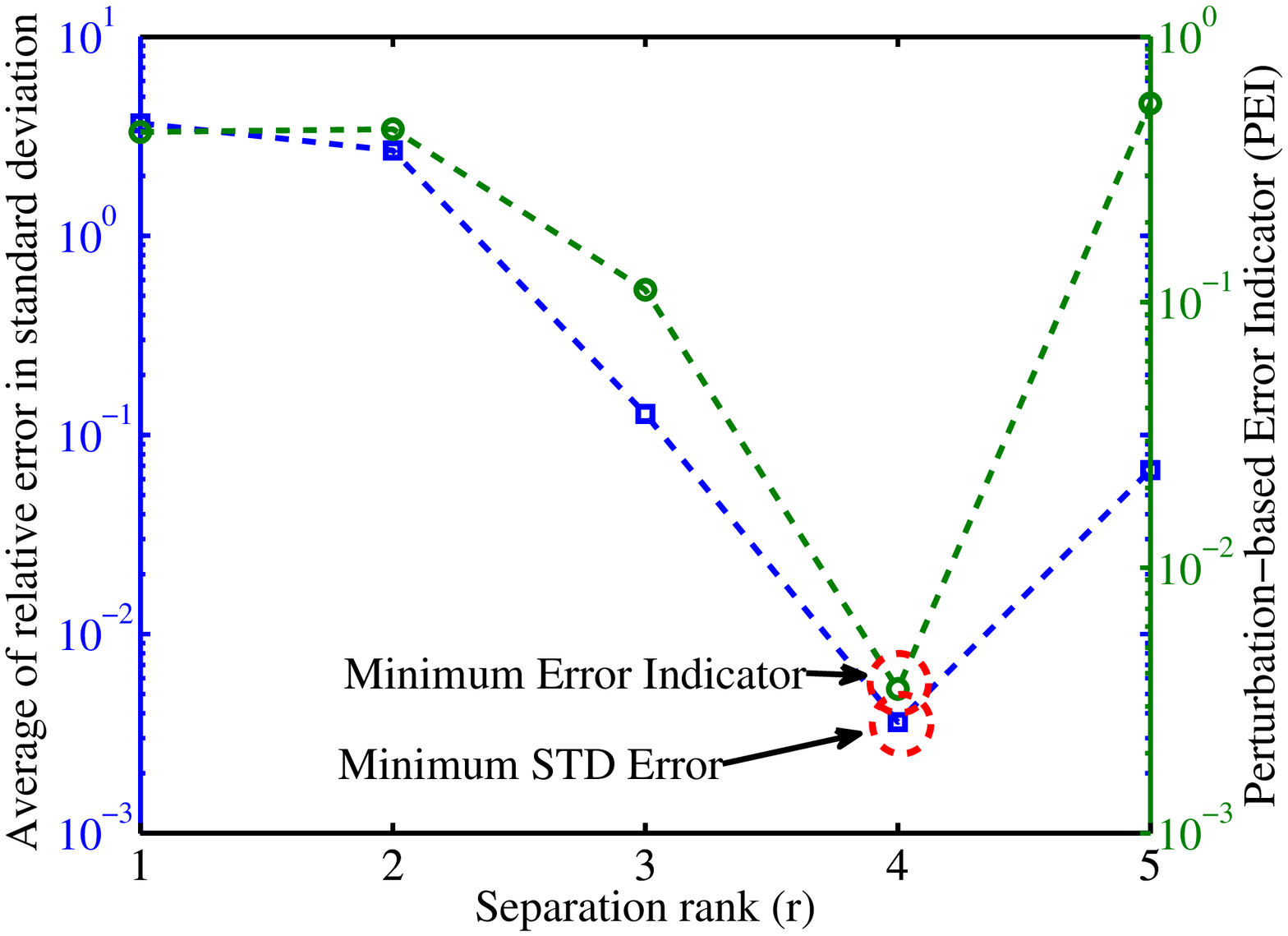}  
      &
      \includegraphics[width=2.6in]{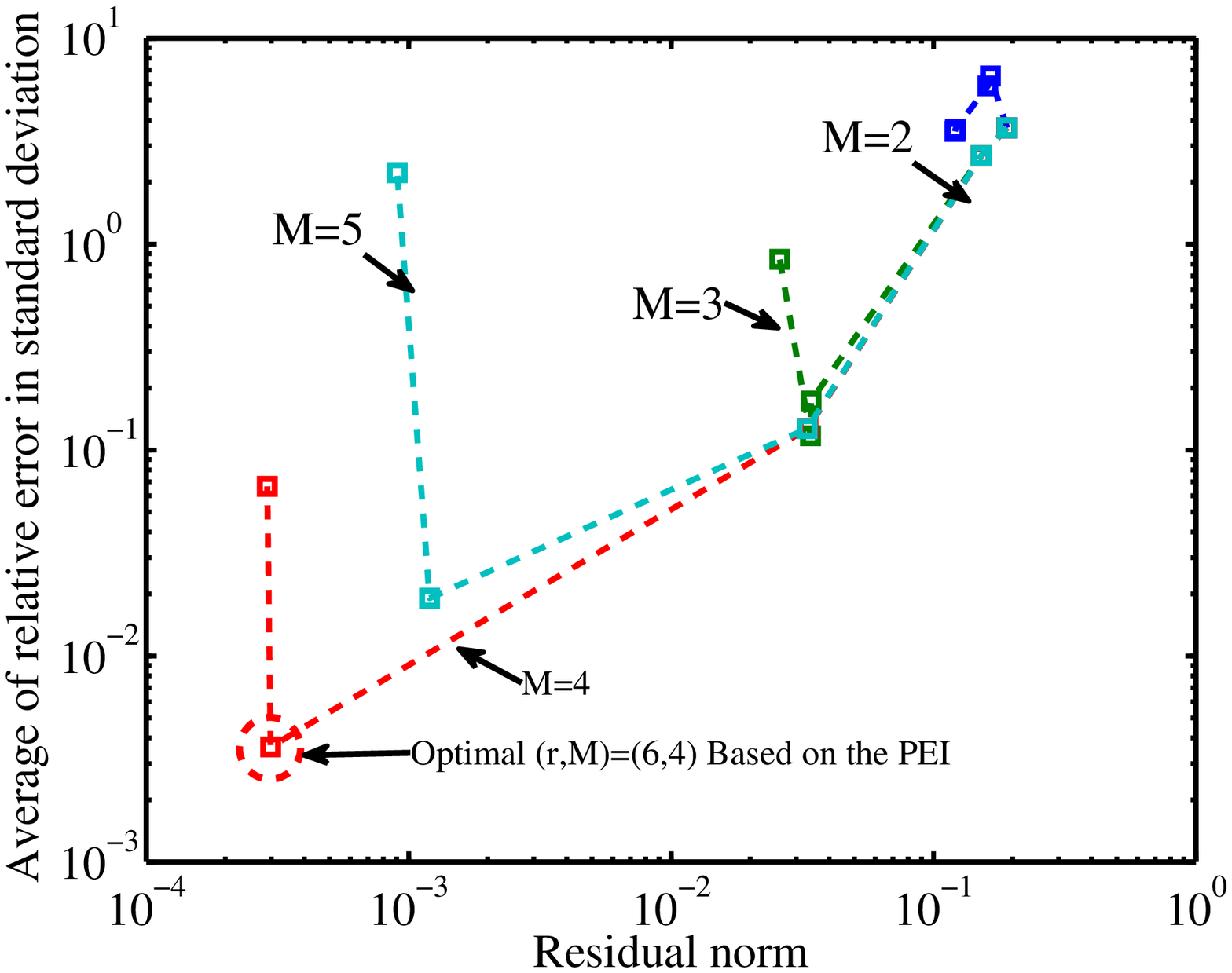} 
      \\
      (a) & (b)           
     \end{tabular}
      \caption{Perturbation based error indicator (PEI) (Section  \ref{subsec: Perturbation based error indicator}) performance in estimating the optimal separation rank and polynomial degree $\left(r,M \right)$. (a) Relative standard deviation error and perturbation based error indicator vs. separation rank  $r$ for the optimal polynomial degree $M$ ; (b) Relative standard deviation error vs. residual norm for different polynomial degree $M$. (Relative error in standard deviation ({\scriptsize$\;\;\;\; \square$} $\hspace{-.73cm} ---$); Perturbation based error indicator (PEI) ($\;\;\;\; \circ $ $\hspace{-.69cm} ---$)).}            
\label{fig:Manufactired_Optimum_rM}       
\end{figure}

\begin{figure}
    \centering
    \begin{tabular}{cc}
            \hspace{-0.5cm}    
      \includegraphics[width=2.7in]{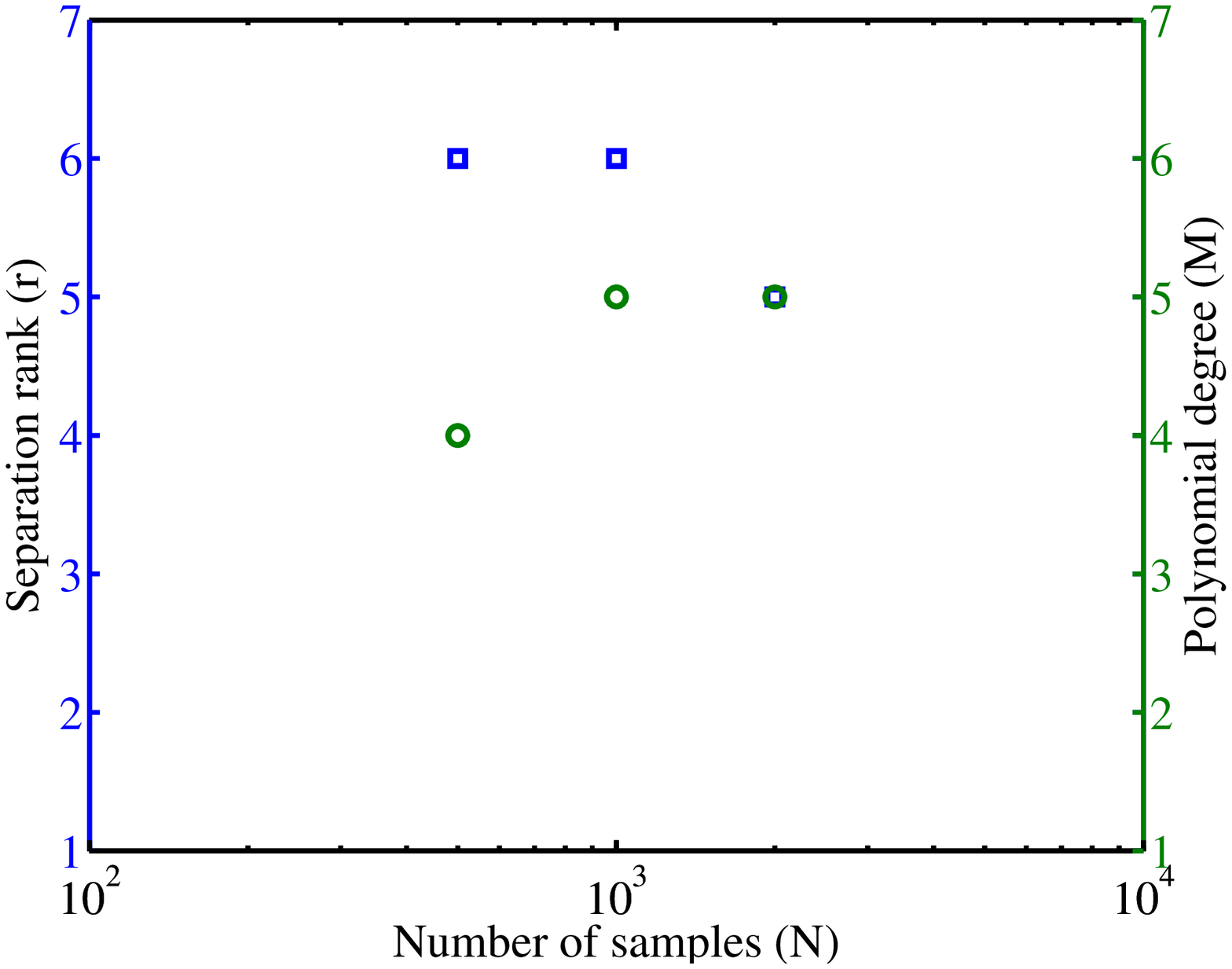}  
      &
      \includegraphics[width=2.9in]{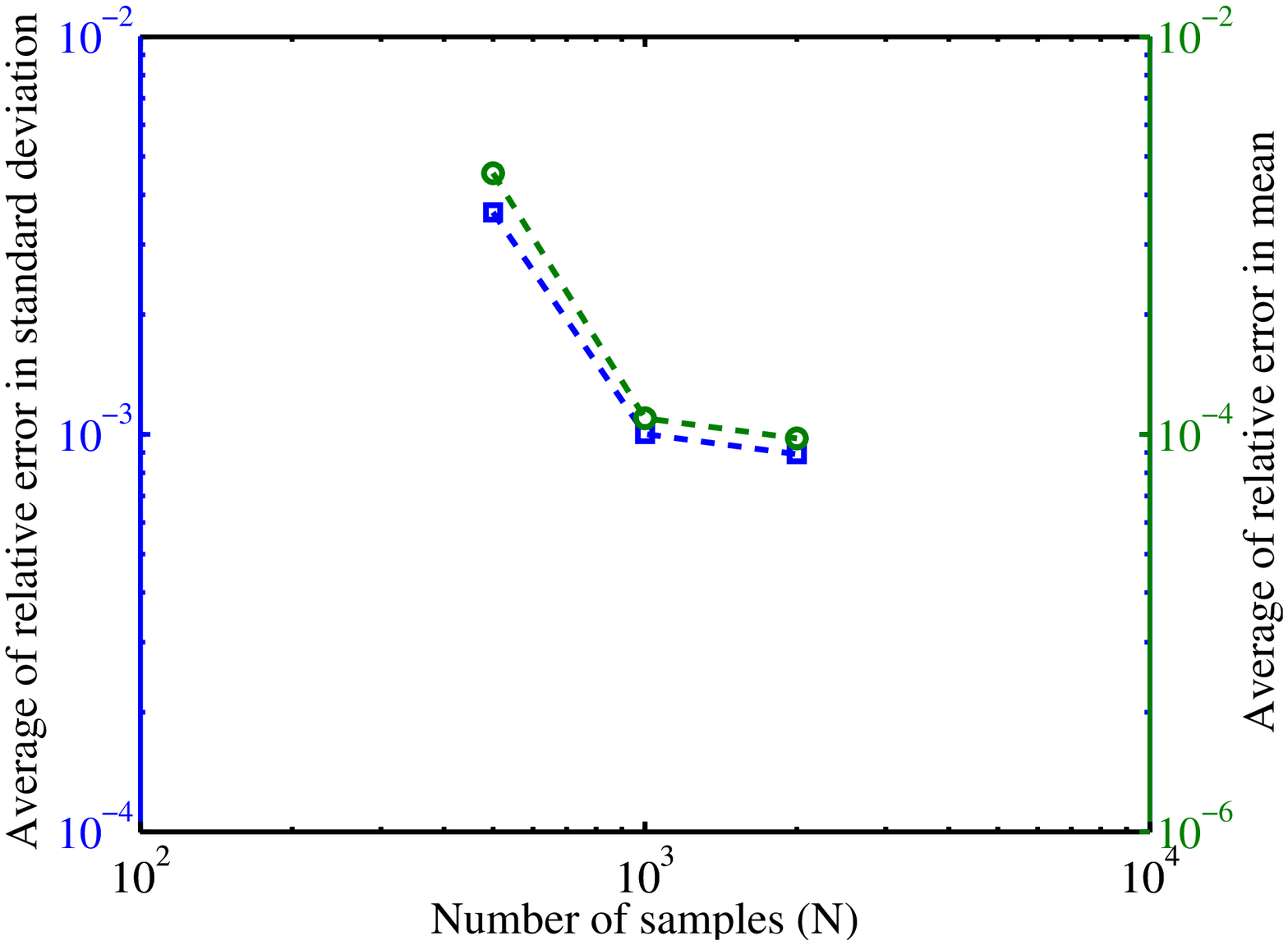} 
      \\
      (a) & (b)
      \\
      \includegraphics[width=2.7in]{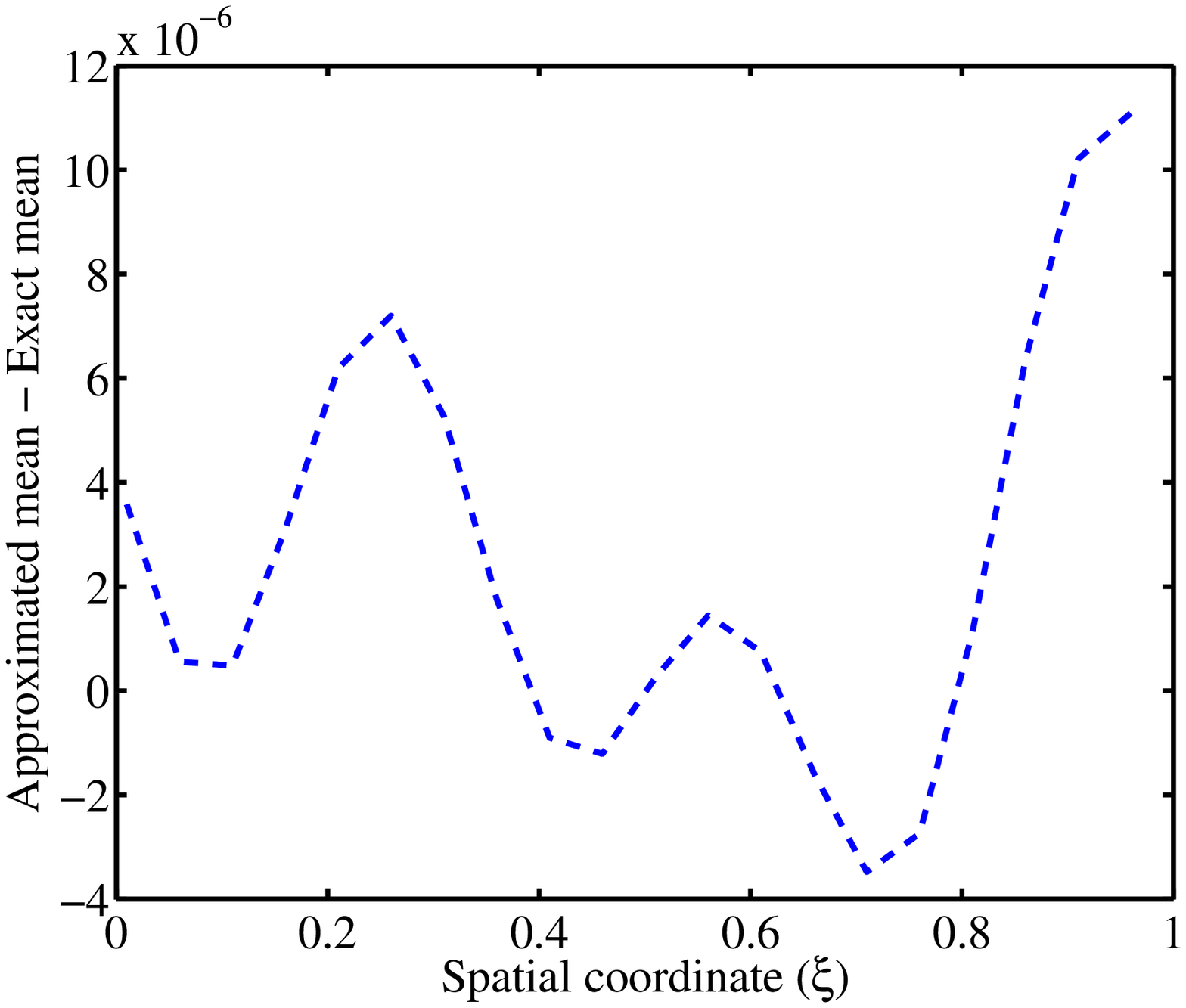}  
      &
      \includegraphics[width=2.7in]{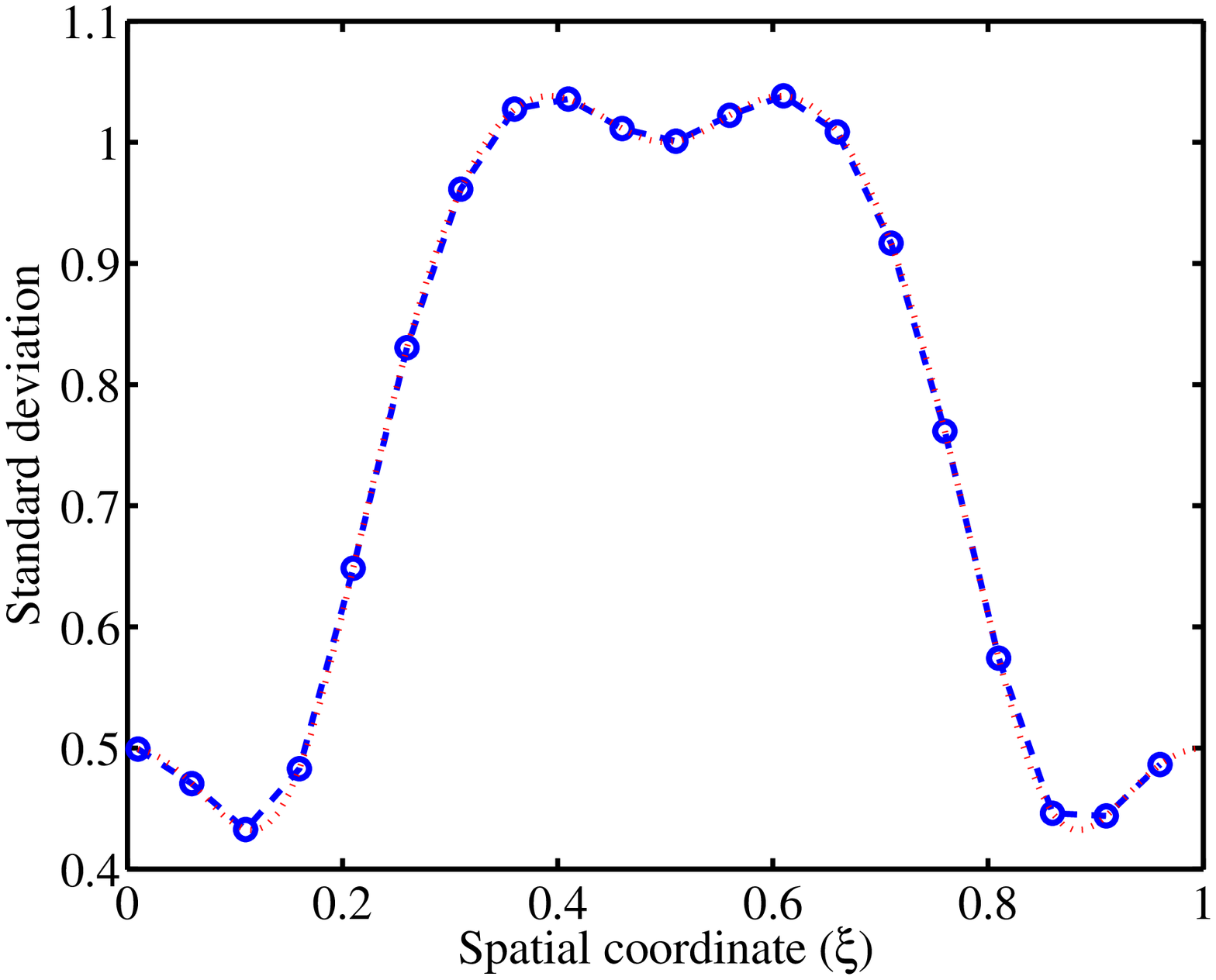}           
      \\
      (c) & (d)   
      
     \end{tabular}
      \caption{ (a) Optimal pair of $(r,M)$ based on PEI vs. number of samples; (b) Relative error in standard deviation and mean vs. number of samples (based on the reference solution); (c) Approximated mean values minus the exact mean equals to $0.55$ vs. spatial variable; and (d) standard deviation of the problem vs. spatial variable. (Separation rank $r$ ({\scriptsize $\square$}); Polynomial degree $M$ ($\circ $);  Relative error in: standard deviation ({\scriptsize$\;\;\;\; \square$} $\hspace{-.73cm} ---$) and mean ($\;\;\;\; \circ $ $\hspace{-.69cm} ---$); Separated representation approximation ($---$); Reference value ($\ldots$)).}             
\label{fig:optimum rM Vs N and std mean}       
       \end{figure}

Here, to verify algorithm $\ref{Algorithm:ALS}$, the following 11-dimensional function is considered:
\begin{equation}
\label{eq:manufactured function}
\bm{u}\left(\xi,\bm{y} \right)=\bm{u}\left(\xi,y_1,\ldots,y_{10} \right) = a_0 + a_1 \sin\left(\pi \xi \right)y_1+a_2 \cos\left(3\pi \xi \right)\left(y_3^2-1\right)
\end{equation}
\begin{equation}
+ \ a_3\sin\left(6\pi \xi\right)\left(y_9^3-3\right)+\bm{\epsilon},
\nonumber
\end{equation}
where $\lbrace y_i\rbrace_{i=1}^{10}$ are independent normal random variables, $\xi$ is a spatial variable, and $\bm{u}$ is a discretized vector of the solution with size $n=20$, i.e., $\bm{u}=\lbrace u_j\rbrace _{j=1}^n$. The coefficients $\lbrace a_i \rbrace_{i=0}^3$ are equal to $\lbrace 0.55, 1,\frac{\sqrt2}{4}$,$\frac{0.1}{\sqrt{6}} \rbrace$. The noise $\bm{\epsilon}$ is a i.i.d standard normal random variable with $0.005$ standard deviation.  The results are shown in  Figures \ref{fig:Manufactired_Optimum_rM} and \ref{fig:optimum rM Vs N and std mean}. 

Figure \ref{fig:Manufactired_Optimum_rM} shows the performance of the perturbation based error indicator (PEI) in estimating the optimal separation rank and polynomial degree for the case $N=500$. In Figure $\ref{fig:Manufactired_Optimum_rM}$.$a$ the PEI shows that the values of pair $\left(r,M \right)=\left(6,4 \right)$ are the optimal model complexities, which correspond to the smallest value of the standard deviation relative error. The PEI functionality for the case $N=500$ is more clear, but for some cases it refers to the pair $\left(r,M \right)$, where the standard deviation error is not necessarily minimal, but it is still acceptable. In figure \ref{fig:Manufactired_Optimum_rM}.$b$, relative standard deviation error with respect to the residual norm is plotted for various $M$. The optimal pair $\left(r,M\right)=\left(6,4\right)$ refers not only to the minimum standard deviation error but also to the smallest residual norm as well. The small separation rank $r=6$ means that this manufactured function can be approximated by a \textit{low-rank} separated model. It also can be shown that larger values of polynomial degree, $(M=5)$, do not necessarily lead to more accurate models. In general, the optimal values of pair $\left(r,M\right)$ are not unique for a problem and might be different for different data sets as well as different numbers of samples. 

Figure \ref{fig:optimum rM Vs N and std mean}.$a$ shows the optimal $\left(r,M\right)$ for different numbers of samples for the same problem. When the number of samples is increased, more information is provided; therefore, the constructed models compute the solutions with higher accuracy. This is shown in figure \ref{fig:optimum rM Vs N and std mean}.$b$, where the error in standard deviation decreases about one order of magnitude by increasing the number of samples from $N=500$ to $N=1000$.  The mean value of the solution is constant and equal to $0.55$. In figure \ref{fig:optimum rM Vs N and std mean}.$c$, the difference of the predicted and exact mean is illustrated with respect to the spatial variable $\xi$. Figure \ref{fig:optimum rM Vs N and std mean}.$d$ shows the standard deviation of the problem based on the separated representation for the case $N=1000$ in comparison to the exact value. It also can be observed from figure \ref{fig:optimum rM Vs N and std mean}.$b$  that the average over the spatial variable of the relative errors in standard deviation and mean are $0.001$ and $0.0001$, respectively.                     

\subsection{Elliptic stochastic equation}
The objective of considering an elliptic stochastic equation is to estimate the diffusion coefficients from available (noisy) observations of the solution field by MCMC, applying the separated representation-based surrogate model.

Here, the forward model,  $\mathcal{A}$, is considered as the following 1D in space elliptic stochastic PDE on the unit interval $\left[\Xi_1,\Xi_2\right]=\left[0,1\right]$ and the continuous domain $\left(\mathcal{D} \subset \mathcal{R}^1\right)$, with Dirichlet boundary conditions on $\partial \mathcal{D}$ \cite{LeMaitre10}:
\begin{eqnarray}
\label{eq:elliptic equation}
\nabla \left(\kappa \left(\xi,\bm{y}\left( w \right) \right) \nabla \bm{u} \left(\xi,w \right) \right)=- f(\xi) \\
u\left(0,\bm{y}\right)=u\left(1,\bm{y}\right)=0,\nonumber
\end{eqnarray}
where $f(\xi)=1$ is assumed as a constant source term . Here, $\nabla \equiv \partial/\partial \xi$ and the PDE is spatially discretized by the finite difference approach with spacing $\Delta \xi=1/1000$ on a uniform grid. The $\kappa$ describes a spatially heterogeneous (diffusion) coefficient as a source of uncertainties that is stochastically discretized by the Karhunen-Loeve  expansion
\begin{equation}
\label{eq:diffusion coefficient}
\kappa\left(\xi,w \right)=\kappa_0 + \exp\left( \sum_{i=1}^d \sqrt{\lambda_i}\phi_i\left(\xi \right) y_i\left(w\right)\right),
\end{equation}
along with an offset $\kappa_0=0.5$. The $\lbrace y_i\rbrace_{i=1}^d$ are independent random variables uniformly distributed on $\left[-1,1\right]$, where $\left(\lambda_i,\phi_i \right)$ are the pairs of eigenvalues and eigenfunctions of the covariance function $c\left(\xi,\xi^\prime \right)$ respectively, i.e.,
\begin{equation}
\label{eq:covariance function}
\int_{\mathcal{D}} c\left(\xi,\xi^\prime \right) \phi\left(\xi^\prime \right) d\xi^\prime = \lambda_i \phi_i\left(\xi\right).
\end{equation}

The $c\left(\xi,\xi^\prime \right)$ is the covariance kernel of the Gaussian process with exponential form as follows:
\begin{equation}
\label{eq:Covariance kernel}
c\left(\xi,\xi^\prime \right) = \sigma^2 exp\left(- \frac{\left( \xi-\xi^\prime \right)^2 }{L_c^2} \right),
\end{equation}
with a prior standard deviation $\sigma=1$ and correlation length $L_c=1/14$. The eigenfunctions, $\phi_i$,  are discretized on the same grid, where the solution field, $\bm{u}$, is discretized. The observations, $\bm{u}$, in equation ($\ref{eq:elliptic equation}$) are obtained at $n=20$ points in $\mathcal{D}$ using a finite element based solver provided in FEniCS. Following algorithm $\ref{Algorithm:ALS}$ with $N=\lbrace 500, 1000, 2000 \rbrace $ samples, the separated representation models are prepared to predict the solution of the elliptic equation ($\ref{eq:elliptic equation}$).

\begin{figure}
    \centering
    \begin{tabular}{cc}
            \hspace{-0.5cm}    
      \includegraphics[width=2.8in]{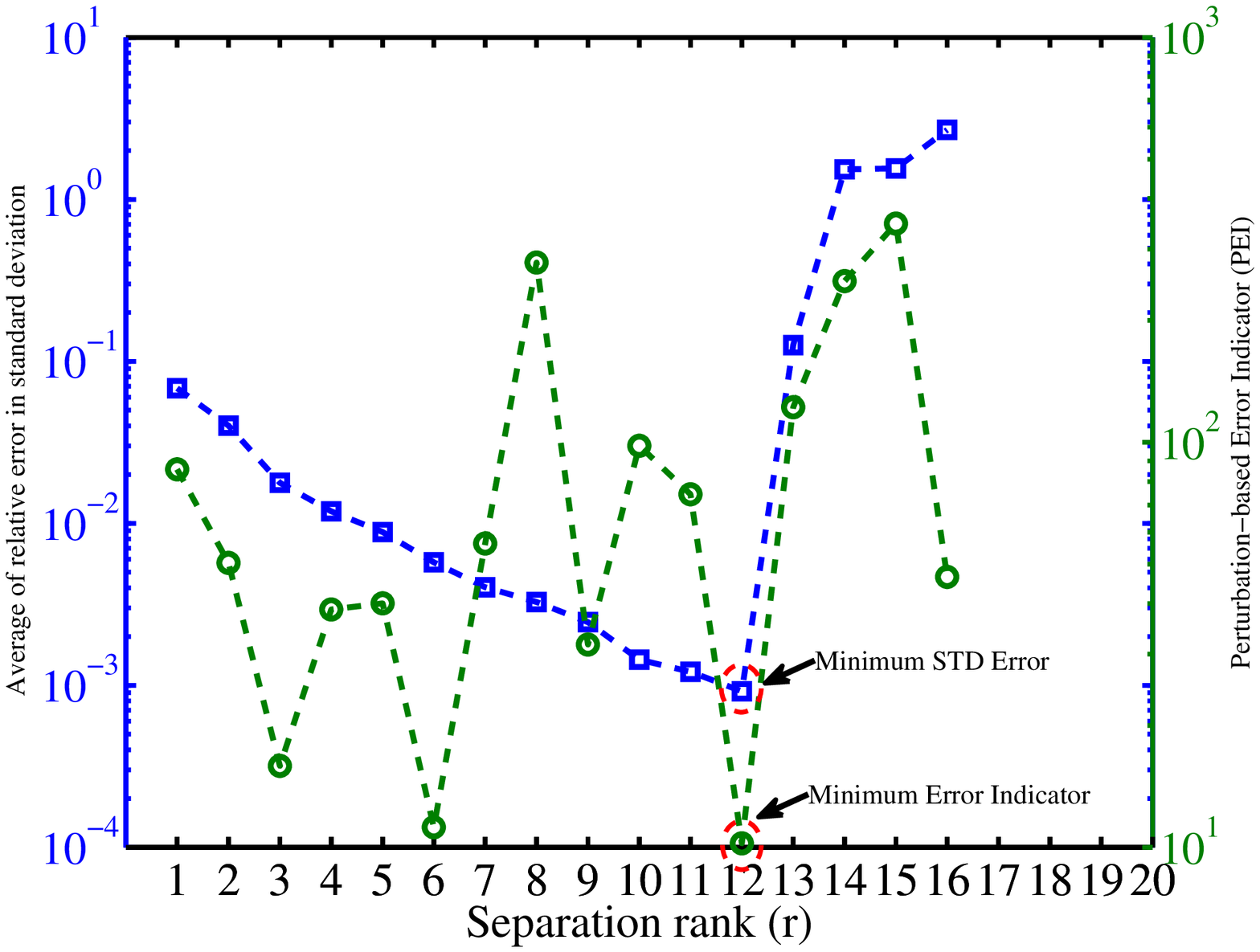}  
      &
      \includegraphics[width=2.7in]{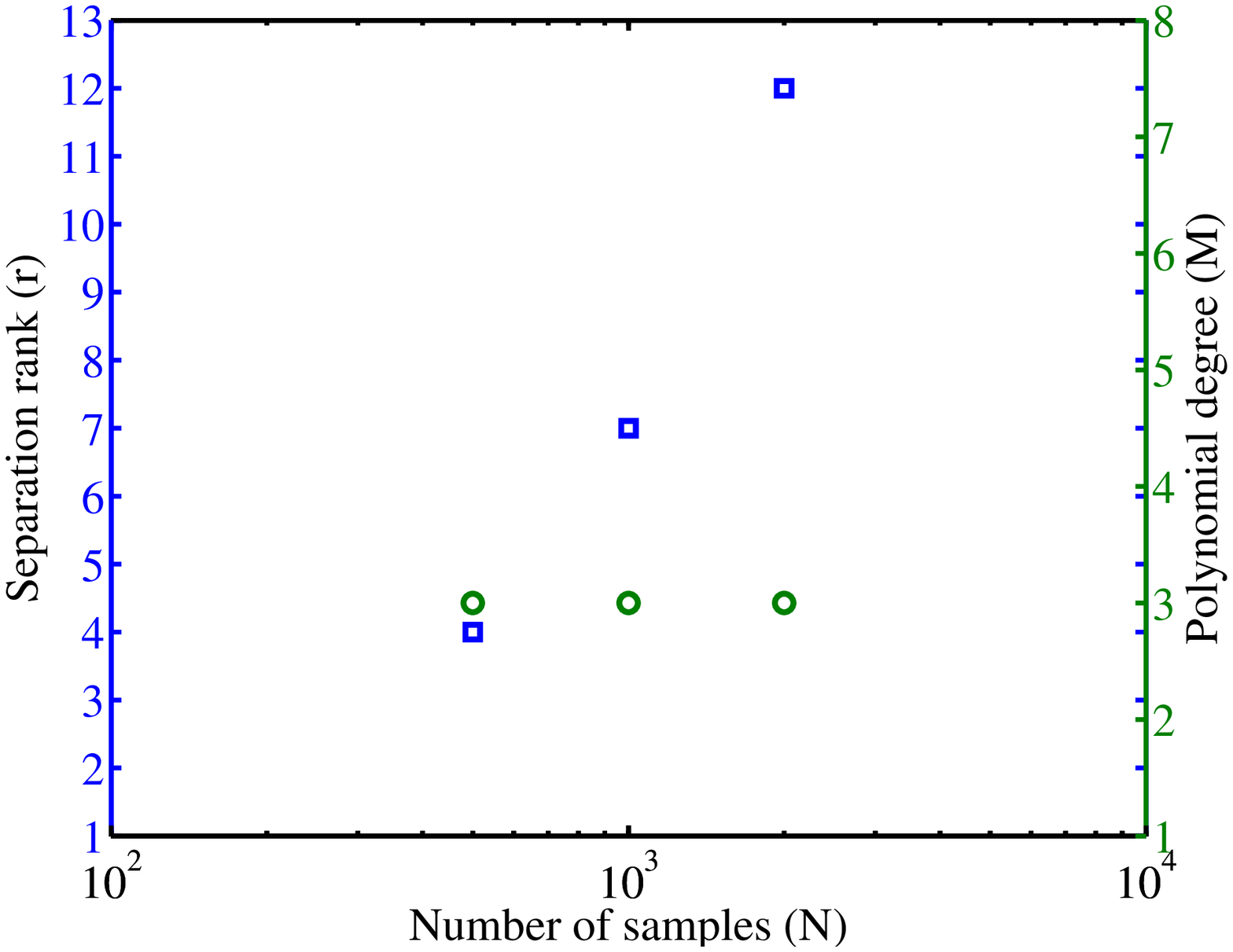} 
      \\
      (a) & (b)            
     \end{tabular}
      \caption{The model performance in estimating the problem's statistics and optimal separation rank and polynomial degree with respect to the number of samples  $N$ (a) Relative standard deviation and mean errors vs. number of samples and (b) Optimal pair of separation rank and polynomial degree vs. number of samples. (Relative error in: standard deviation ({\scriptsize$\;\;\;\; \square$} $\hspace{-.73cm} ---$) and mean ($\;\;\;\; \circ $ $\hspace{-.69cm} ---$); Optimal separation rank ({\scriptsize$\square$}); Optimal polynomial degree ($\circ$)).}            
\label{fig:optimal rM}       
       \end{figure}

 To find the optimal pair of $\left(r,M \right)$, the minimum value of PEI is found. Figure $\ref{fig:optimal rM}$.$a$ shows the standard deviation error and PEI for the case of $N=2000$ and $M=3$. In this case, the minimum value of PEI corresponds to a separation rank of $12$ and polynomial degree of $3$, where the standard deviation error is also minimal. The optimal pair of $\left(r,M\right)$ is obtained with the same analysis for different number of samples. As can be seen in figure $\ref{fig:optimal rM}$.$b$, there is no unique separation rank and polynomial degree for the problem, and these values depend on the number of samples and the data set, e.g., the optimal pairs for $N= 1000$ and $2000$ are equal to $\left(7,3\right)$ and $\left(12,3\right)$, respectively. 

\begin{figure}
    \centering
    \begin{tabular}{cc}
            \hspace{-0.5cm}    
      \includegraphics[width=2.7in]{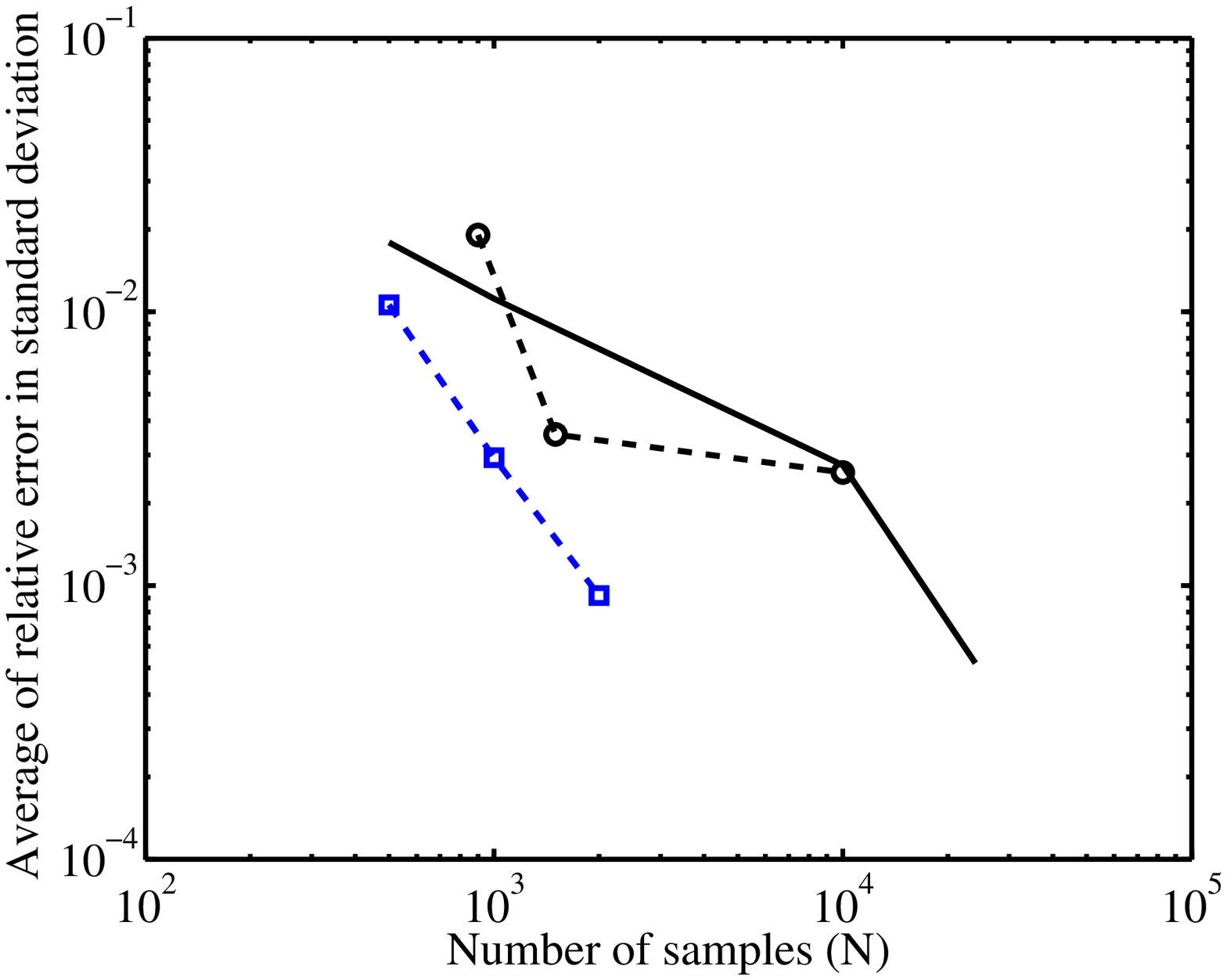}  
      &
      \includegraphics[width=2.7in]{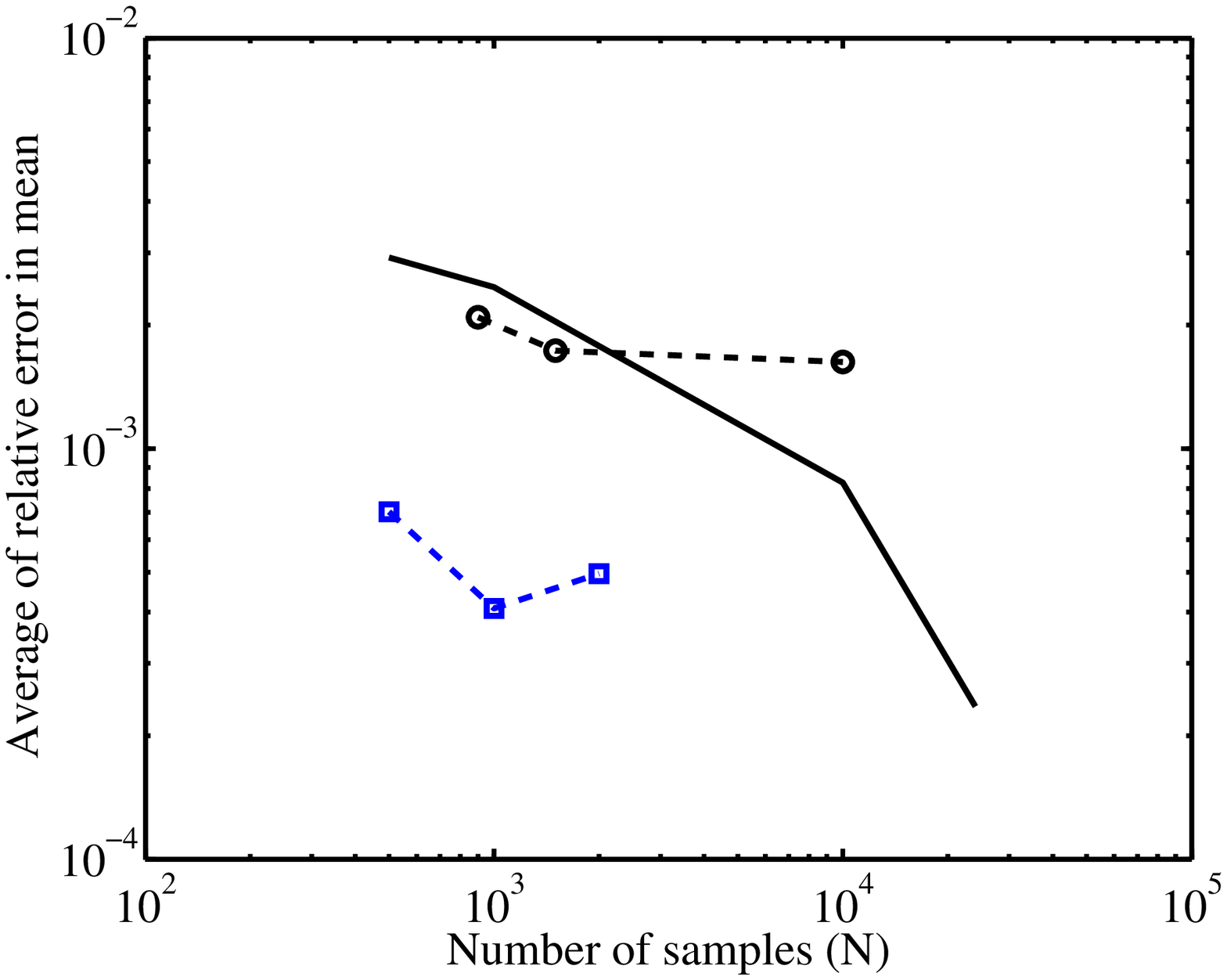} 
      \\
      (a) & (b)            
     \end{tabular}
      \caption{Comparison of the average relative errors in standard deviation and mean of the solution for the separated representation, regression and Monte Carlo: (a) Relative standard deviation error; (b) Relative mean error (Separated representation ({\scriptsize$\;\;\;\; \square$} $\hspace{-.73cm} ---$); Regression ($\;\;\;\; \circ $ $\hspace{-.69cm} ---$); Monte Carlo $\left(\line(1,0){10} \right)$ ) }            
\label{fig:comparison std and mean rel ative errors }       
       \end{figure}

\begin{figure}
    \centering
    \begin{tabular}{cc}
            \hspace{-0.5cm}    
      \includegraphics[width=2.7in]{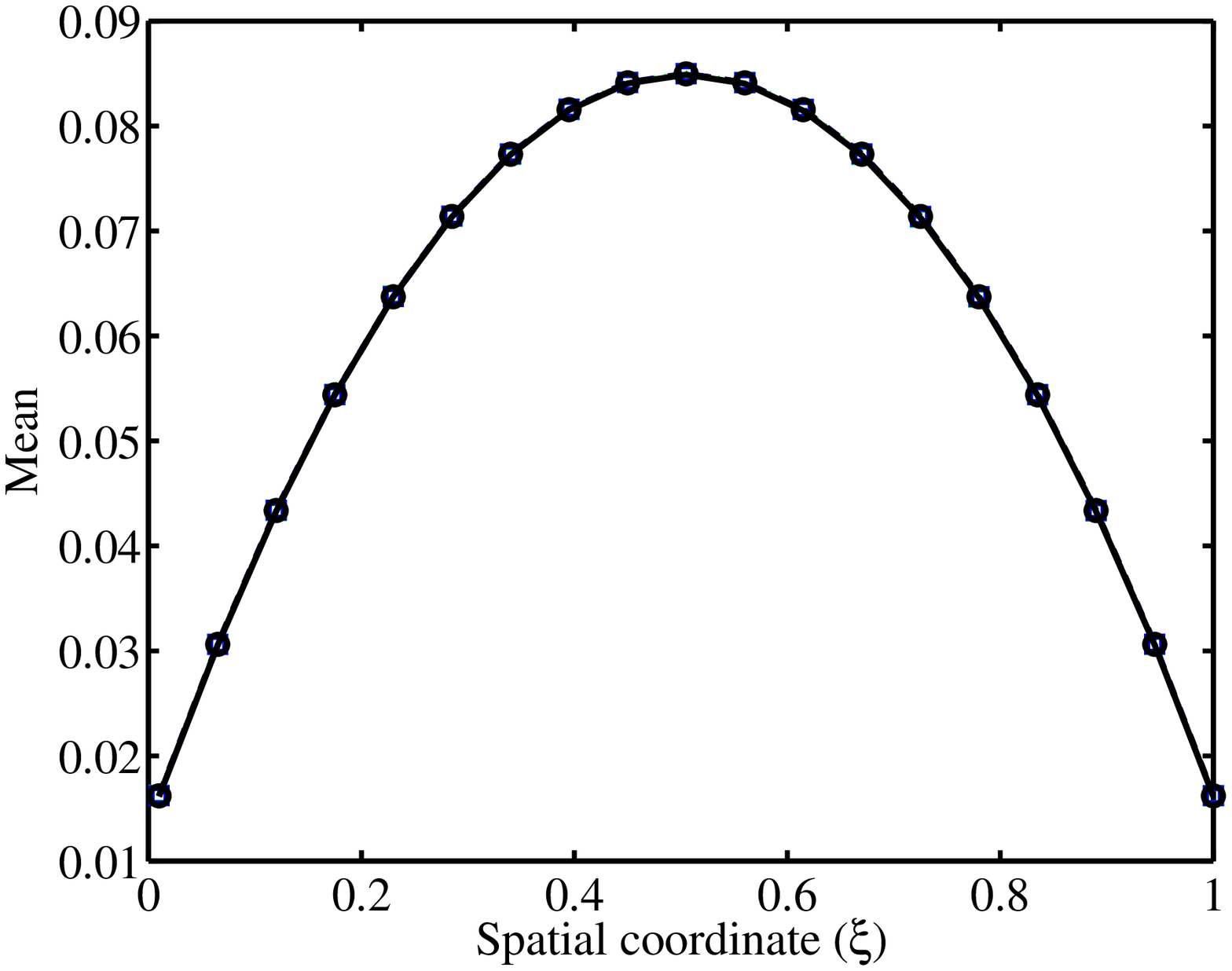}  
      &
      \includegraphics[width=2.7in]{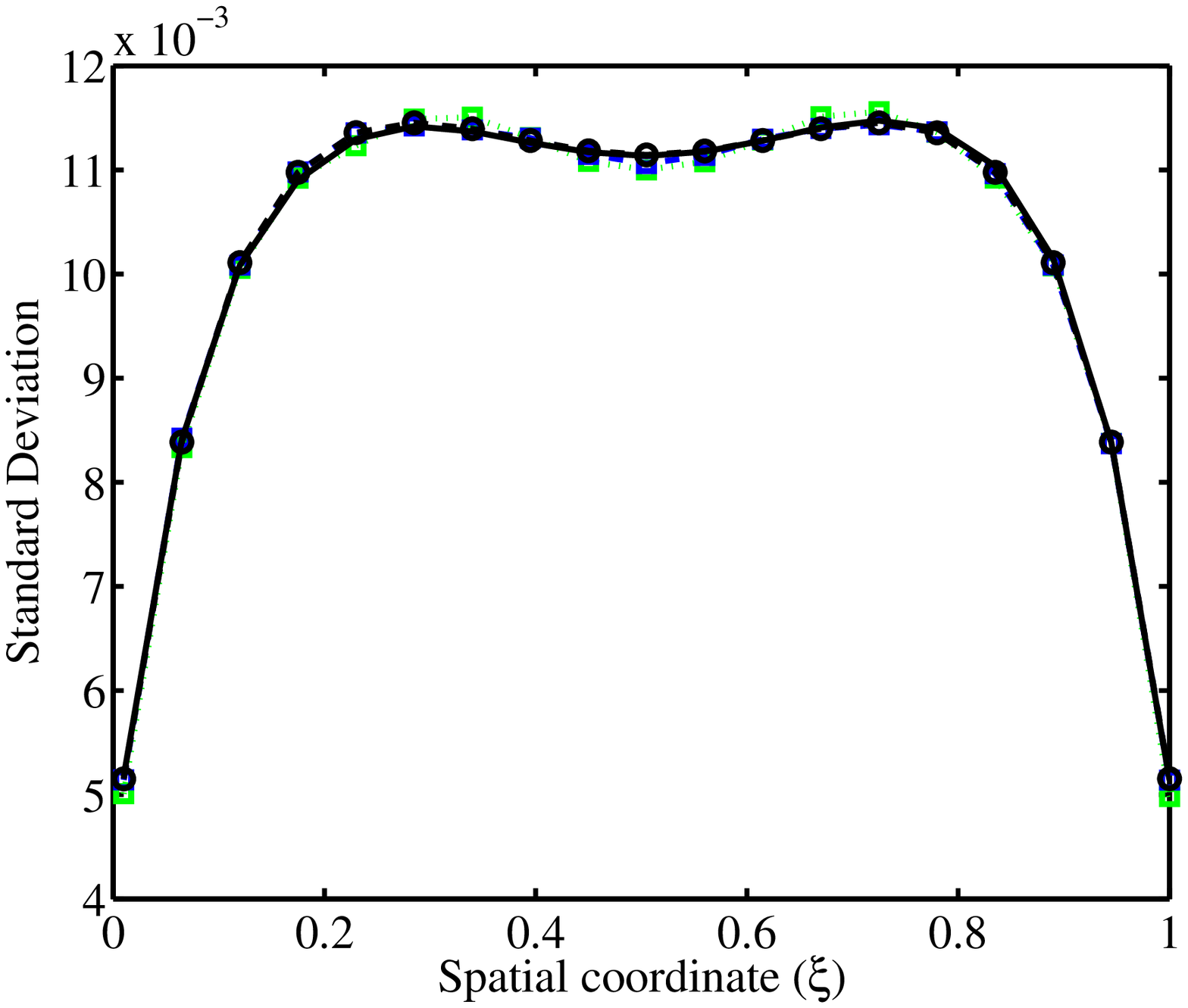} 
      \\
      (a) & (b) 
           \\           
     \end{tabular}
      \caption{Statistics of the elliptic PDE solution as a function of spatial variable $(\xi)$ (a) Mean; (b) Standard deviation; (Estimated based on: Separated representation with: $N=1000$ ({\scriptsize$\;\;\;\; \square$} $\hspace{-.73cm} \ldots\ldots$) and $N=2000$ ({\scriptsize$\;\;\;\; \square$} $\hspace{-.73cm} ---$); Regression with $N=25000$ ($\;\;\;\; \circ $ $\hspace{-.69cm} ---$); Monte Carlo with $N=25000$ $\left(\line(1,0){10} \right)$ .}            
\label{fig:Elliptic mean and std}       
       \end{figure}

Figure $\ref{fig:comparison std and mean rel ative errors }$ illustrates the average relative errors in standard deviation and mean with respect to the number of samples. The average is over the physical variable of size $n$. The reference statistics are obtained by the Monte Carlo (MC) approach with $N=25000$ samples to compare the predicted results with. It can be seen that the separated representation model with $N=2000$ predicts an average of mean and standard deviation relative errors equal to $9\times 10^{-4}$ and $4\times 10^{-4}$, respectively. In figure $\ref{fig:comparison std and mean rel ative errors }$, the accuracy of the separated representation model is compared to regression and MC approaches. It is shown that to get the same accuracy the separated model needs fewer samples by one order of magnitude. The problem statistics which are illustrated in figures $\ref{fig:Elliptic mean and std}$.$a$ and $\ref{fig:Elliptic mean and std}$.$b$ are compared to the regression results and the reference values. 

Up to this point, an accurate surrogate separated representation model to approximate the solution of the equation ($\ref{eq:elliptic equation}$) has been provided and validated. In the following, it will be shown how this surrogate model can be used in inverse problems. Based on the true value of diffusion coefficient $\kappa$, equation ($\ref{eq:elliptic equation}$) is solved to obtain the exact value of the solution, $\bm{u}_{exact}$. The true $\kappa$ is obtained by generating uniform independent random variables, $\lbrace y_i \rbrace_{i=1}^d$, and computing equation ($\ref{eq:diffusion coefficient}$). In the inverse problem context, one needs noisy observations, e.g., experimental results. To do so, an i.i.d standard normal noise vector is added to the $\textit{exact}$ solution, i.e., $\bm{u}_{noisy}=\bm{u}_{exact}+\bm{\varepsilon}$, where $\varepsilon_j\sim N\left(0,\sigma_{noise}\right),\ j=1,\ldots,n$.

\begin{figure}
    \centering
    \begin{tabular}{cc}
            \hspace{-0.5cm}    
      \includegraphics[width=2.6in]{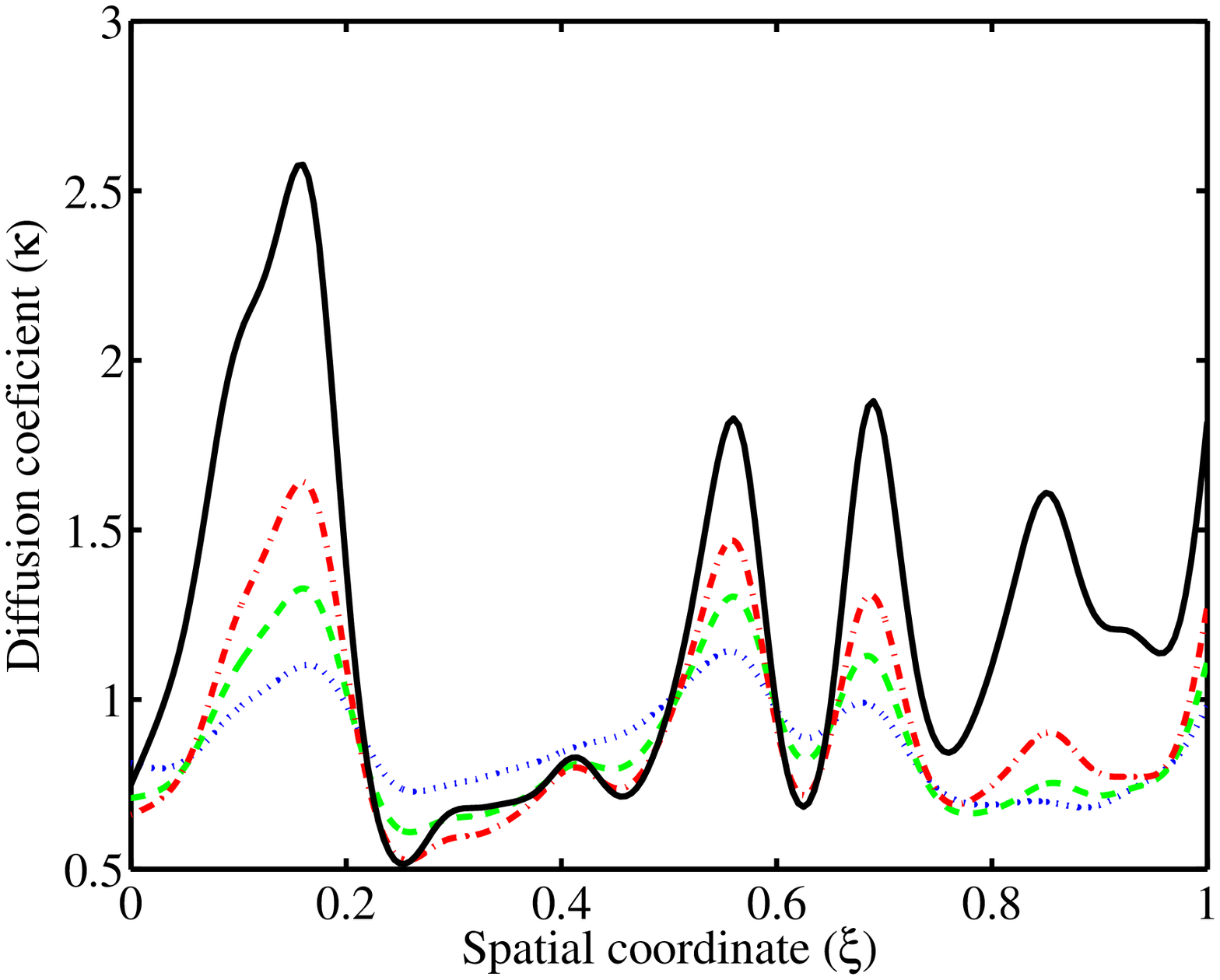}  
      &
      \includegraphics[width=2.7in]{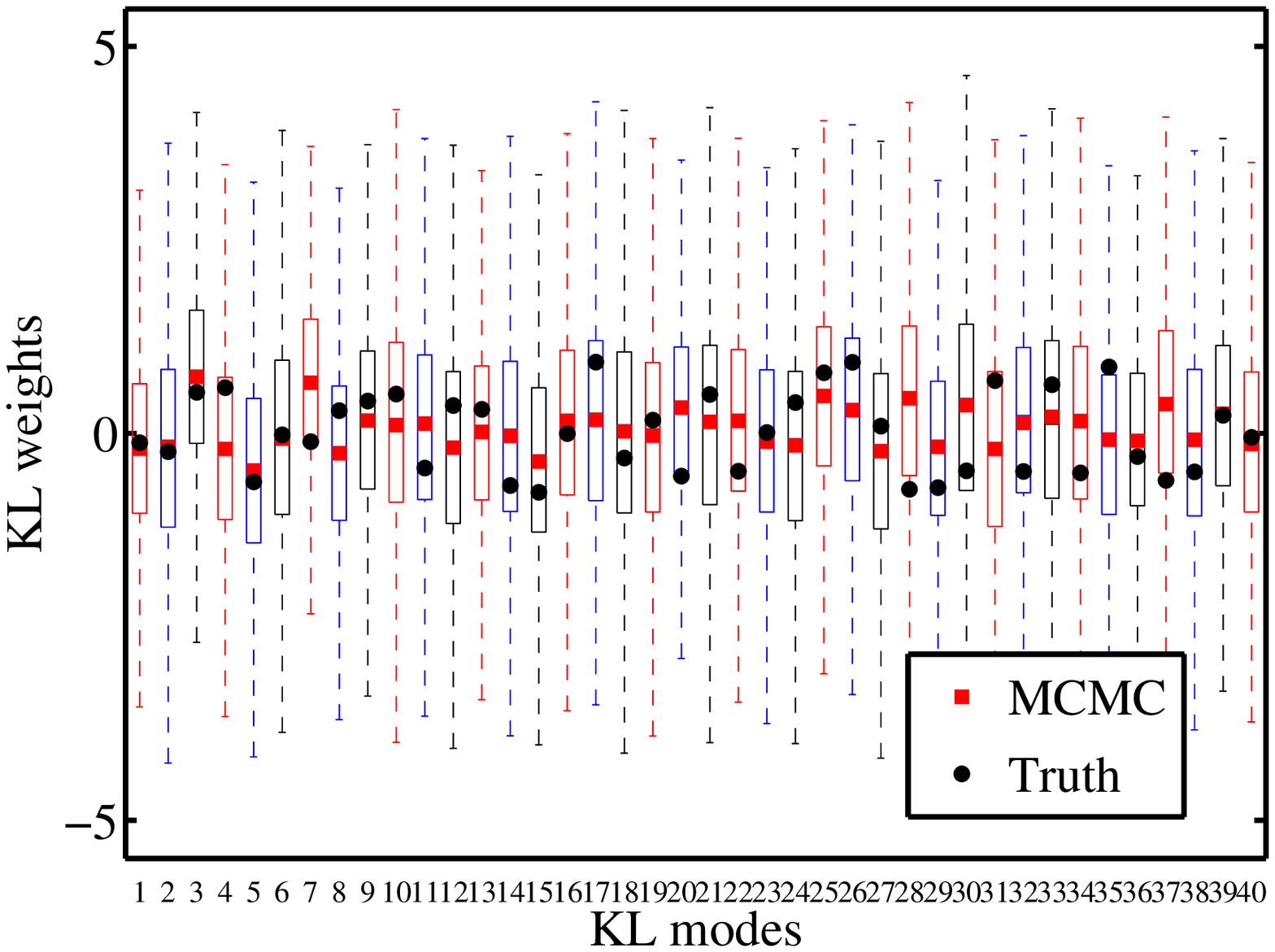} 
      \\
      (a) & (b)            
     \end{tabular}
      \caption{MCMC results for one-dimensional elliptic PDE. (a) Diffusion coefficient obtained with posterior realizations vs. the spatial variable. ($\sigma_p = 0.3$ ($\;\;\;\;\;\;  $ $\hspace{-.69cm} .-.-.-$),  $\sigma_p = 0.6$ ($\;\;\;\;\; $ $\hspace{-.69cm} ---$),  $\sigma_p = 1$ ($\;\;\;\;\;\; $ $\hspace{-.69cm} \ldots$), and truth (solid line). (b) Posterior boxplot obtained with the MCMC. (Posterior mean {\scriptsize$\left( \square \right)$} and the true value of KL weights $\left(\circ\right)$.}            
\label{fig:MCMC results}       
       \end{figure}

Here, one case where $\sigma_{noise}=0.05$ and $n=20$ is considered for the $41$-dimensional elliptic PDE ($\ref{eq:elliptic equation}$). To decrease the computational cost of repeated evaluation of the forward model in Bayesian inference approaches, the separated representation surrogate model is used to approximate the elliptic PDE solution, $\bm{u}$. The results are shown in figures $\ref{fig:MCMC results}$.$a$ and $\ref{fig:MCMC results}$.$b$. Figure $\ref{fig:MCMC results}$.$a$ illustrates the diffusion coefficients based on the posterior realizations computed by DRAM \citep{Heikki06}. Regarding the prior information, the KL modes in equation (\ref{eq:diffusion coefficient}) are uniformly distributed; however, a Gaussian prior distribution with $prior_{mean}$ and $prior_{std}$ is considered and transformed to uniform distribution on $\left[-1,1\right]$ by 
\begin{equation}
\label{eq:normal to uniform distribution}
y = erf\left( \left(\theta - prior_{mean} \right) / \left( \sqrt{2} \times prior_{std} \right) \right) \sim U\left(-1,1 \right).
\end{equation}

Here, $\theta$ is a normal random variable $\theta \sim N(prior_{mean},prior_{std})$ and $erf$ is the error function.  
Figure ($\ref{fig:MCMC results}$.$a$) shows the results of inverse modelling for three cases with prior Gaussian distributions and the same $prior_{mean}$, but different $prior_{std}$. The $prior_{mean}$ value is approximated by the method described in \cite{Jeffrey98}, which minimizes the square root of the differences between the approximation and the exact value of the solution. The prior standard deviations are equal to $\lbrace 0.3, 0.6, 1\rbrace$. The solid line represents the true value of the diffusion coefficients used to generate the data and the remaining lines represent the DRAM simulations with various prior standard deviations. The results are obtained using $10^6$ DRAM samples, disregarding $2.5 \times 10^5$ of them as burn-in samples. As expected, due to the ill-conditioning of the problem the non-smooth part of the diffusion field is not easy to reconstruct. However, by decreasing the prior-standard deviation, which defines more narrower bound around the $prior_{mean}$ for the realizations, more features of the diffusion field can be captured.  

Figure $\ref{fig:MCMC results}$.$b$ illustrates a boxplot of KL mode weights superimposed with the posterior mean of realizations obtained with DRAM. In each boxplot a central box represents the center ($50 \%$) of the posterior, a central line indicates the median, upper and lower lines represent the $25 \%$ and $75 \%$ quantile of the posterior, and two vertical lines, or whiskers indicate the entire range of the posterior outside the central box. The exact value of the realizations which were used to generate the data are also shown in this figure. It is shown in the figure $\ref{fig:MCMC results}$.$b$ that the lower modes are approximated more accurately than the higher modes. Because the elliptic operator in equation ($\ref{eq:elliptic equation}$) smooths the higher index-modes and consequently they are rougher and more difficult to reveal by MCMC than the lower index-modes.

\subsection{Cavity Flow}
\label{subsec:Cavity flow}
To investigate the performance of the algorithm in a more challenging problem, a square 2D in space cavity flow is considered, which is filled with a Newtonian fluid of density $\rho_c$, molecular viscosity $\mu_c$, and thermal conductivity $\kappa_c$. The right and left vertical walls are maintained at $T_h$ and $T_c$ temperature, respectively, where $T_h>T_c$. The two horizontal walls are assumed to be adiabatic. Furthermore, the reference temperature and the temperature difference are defined as:
\begin{equation}
\label{eq:ref temp}
T_{ref} = \left(T_h+T_c \right)/2,
\end{equation}
and 
\begin{equation}
\label{eq:temp difference}
\Delta T_{ref} = T_h - T_c,
\end{equation}
where the two temperature boundary conditions are $T_h=-T_c=-1/2$. Using the Boussinesq approximation the normalized governing equations can be derived as follows:
\begin{equation}
\label{eq:cavity1}
\frac{\partial \mathbf{u}}{\partial t} + \mathbf{u}\cdot \nabla \mathbf{u} = - \nabla p + \frac{Pr}{\sqrt{Ra}} \nabla ^2\mathbf{u}+Pr\ T\ \mathbf{y},
\end{equation}

\begin{equation}
\label{eq:cavity2}
\nabla \cdot \mathbf{u}= 0,
\end{equation}

\begin{equation}
\label{eq:cavity3}
\frac{\partial T}{\partial t} + \nabla \cdot \left( \mathbf{u} T \right) = \frac{1}{\sqrt{Ra}} \nabla ^2 T,
\end{equation}
where $\mathbf{u}$ is velocity, $t$ is time, $p$ is pressure, and $T$ is normalized temperature such that $T \equiv \left(T-T_{ref}\right)/T_{ref}$. The $Pr$ and $Ra$ are Prandtl and Rayleigh numbers, which are equal to 0.71 and $10^6$, respectively, which values lead to a steady laminar circulating flow. The equations are discretized on a $1000 \times 1000$ grid. The temperature on the cold wall $(\xi_1=1)$ is expressed as: 
\begin{equation}
\label{eq:cold wall temp}
T\left(\xi_1=1,\xi_2 \right) = T_c +T'(\xi_2).
\end{equation}  

The mean temperature along the cold wall, $\left\langle T\left(\xi_1=1,\xi_2 \right) \right\rangle$, is assumed  to be independent of $\xi_2$ and equal to $T_c=-0.5$. Here, the effect of the cold wall temperature fluctuations $T'(\xi_2)$ on the temperature field is studied. These fluctuations are approximated by a truncated KL expansion with $d=20$ terms as follows:
\begin{equation}
\label{eq:cold wall temp fluctuation}
T'\left(\xi_2\right)=\sum_{i=1}^d \sqrt{\lambda_i}\phi_i\left(\xi_2 \right) y_i\left(w\right),
\end{equation}  
where $\left\lbrace y_i \right\rbrace_{i=1}^d$ are independent input random variables with Gaussian distributions. $\lambda_i$, the eigenvalues, are computed by 
\begin{equation}
\label{eq:Covariance kernel}
\lambda_i = \sigma_c^2 \frac{2L_c }{1+\left(\omega_i L_c\right)^2} \ ,
\end{equation}
in which $\omega_i$ are positive roots of the characteristic equation
\begin{equation}
\label{eq:Characteristic equation}
\left[1-L_c \ \omega_i \tan(\omega_i/2) \right] \left[L_c \ \omega_i+\tan(\omega_i/2) \right]=0.
\end{equation}

The $L_c$ and $\sigma_c$ are $1/21$ and $11/100$, respectively.
In equation (\ref{eq:cold wall temp fluctuation}), $\phi_i$ are eigenfunctions which are given by
\begin{equation}
\label{eq:Eigenfunctions}
\phi_i\left(\xi_2\right) = \left\lbrace \begin{tabular}{c c}
$\frac{\cos \left[\omega_i \left(\xi_2-0.5 \right) \right]}{\sqrt{0.5 + \frac{\sin\left(\omega_i \right)}{2\omega_i}}}$ \\
			 $\frac{\sin \left[\omega_i \left(\xi_2-0.5 \right) \right]}{\sqrt{0.5 - \frac{\sin\left(\omega_i \right)}{2\omega_i}}}$ 
			\end{tabular} \right. .
\end{equation}

\begin{figure}  
\centering
\begin{tabular}{c}
\hspace{-0.5cm}    
\includegraphics[width=3in]{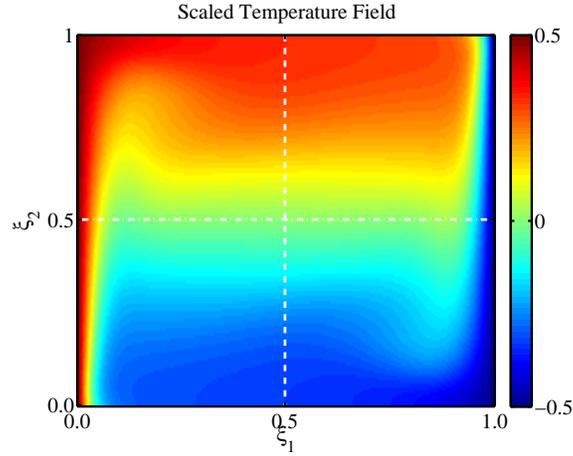}  
\end{tabular}
\caption{Scaled temperature field of a square 2D cavity flow and assigned vertical line $\xi_1=0.5$ ($.-.-.-$) and horizontal line $\xi_2=0.5$ ($---$) for separated representation approximations.}            
\label{fig:Cavity flow field}       
\end{figure}

\begin{figure}
    \centering
    \begin{tabular}{cc}
            \hspace{-0.5cm}    
      \includegraphics[width=2.7in]{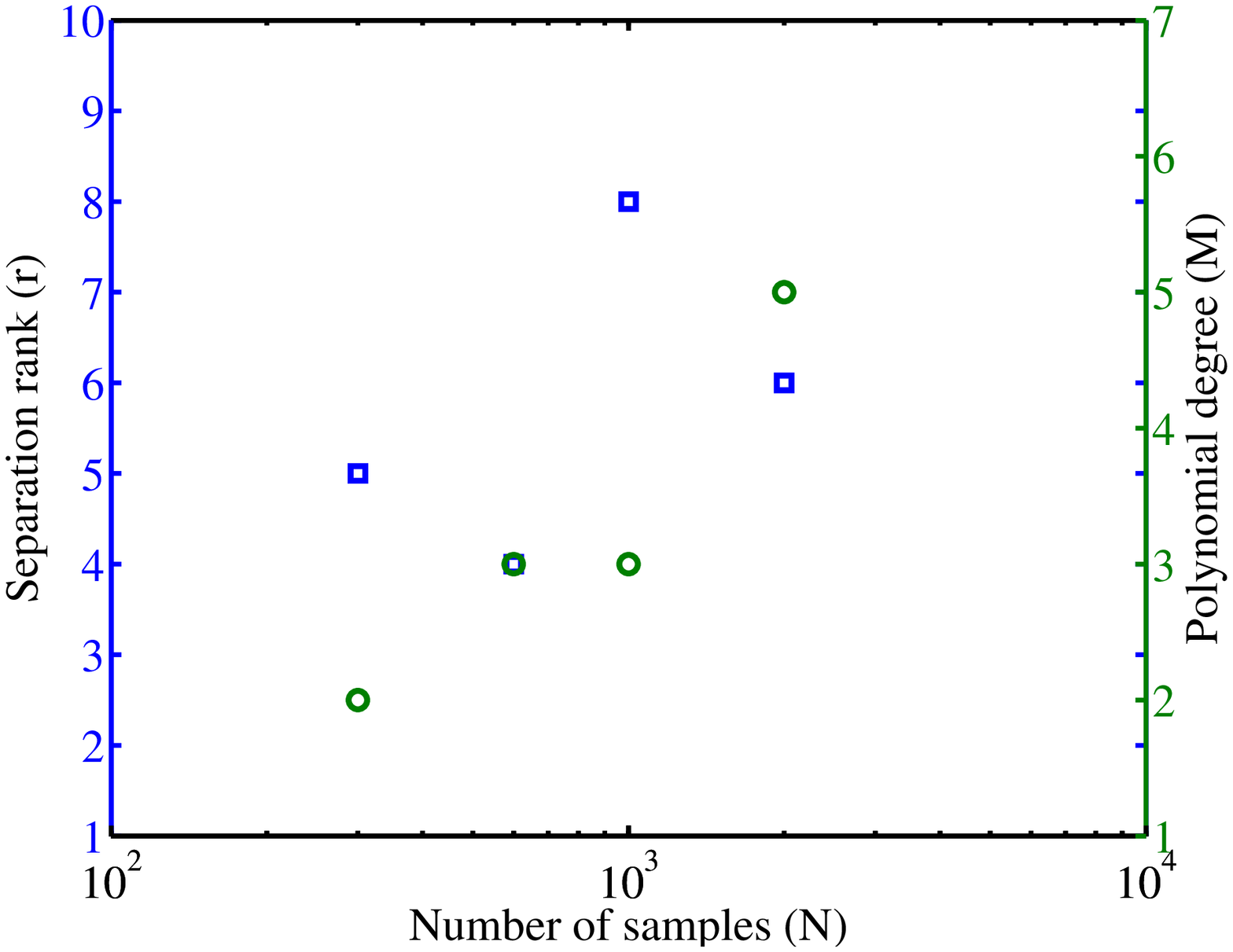}  
      &
      \includegraphics[width=2.7in]{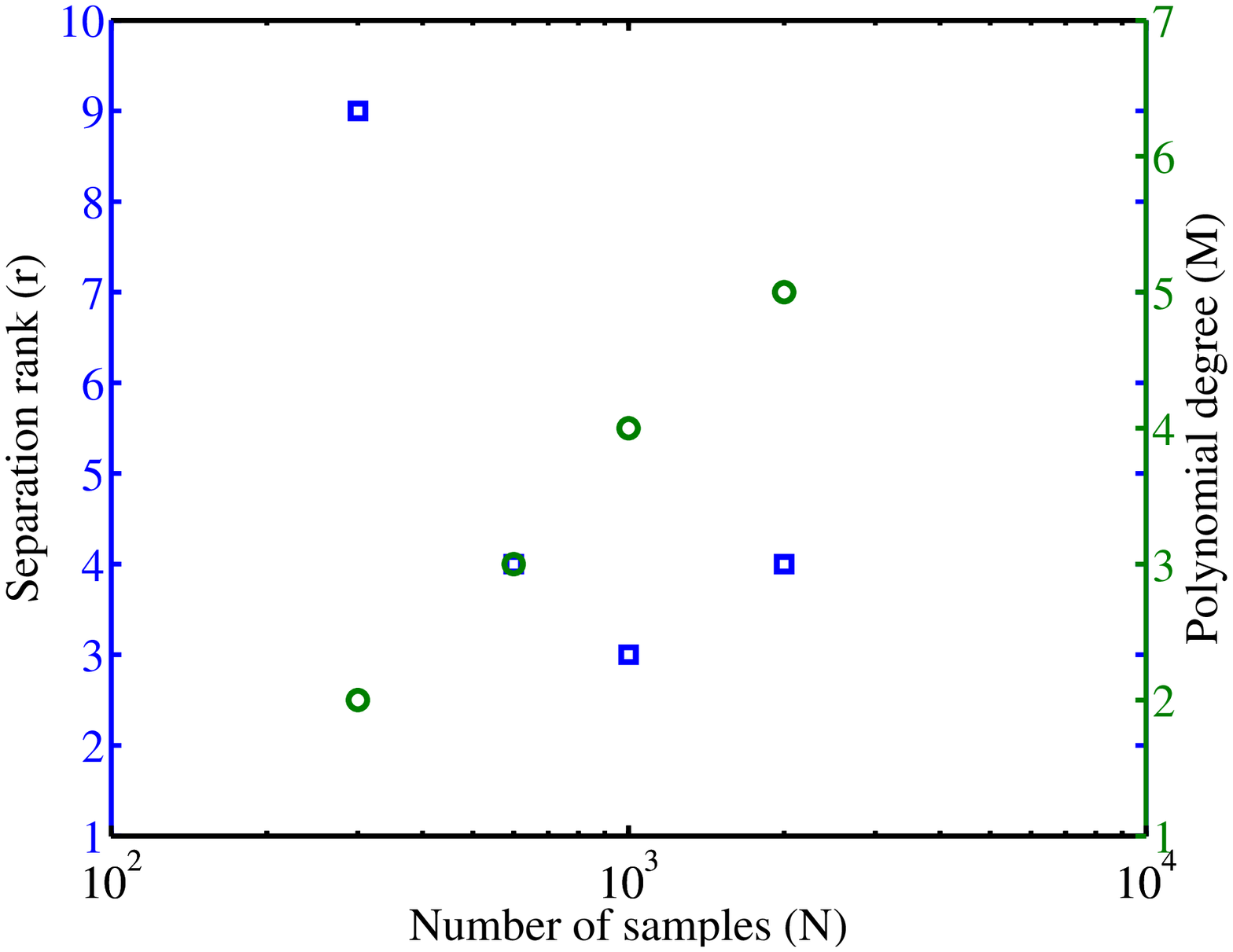} 
      \\
      (a) & (b)            
     \end{tabular}
      \caption{Optimal values of separation rank $r$ and spectral polynomial degree $M$ obtained by following algorithm \ref{Algorithm:ALS} for the lines: (a) $\xi_1=0.5$ and (b) $\xi_2=0.5$. (Separation rank $r$ ({\scriptsize $\square$}); Polynomial degree $M$ ($\circ $)).}            
\label{fig:Optimal rM for cavity flow}       
       \end{figure}

\begin{figure}
    \centering
    \begin{tabular}{cc}
            \hspace{-0.5cm}    
      \includegraphics[width=2.7in]{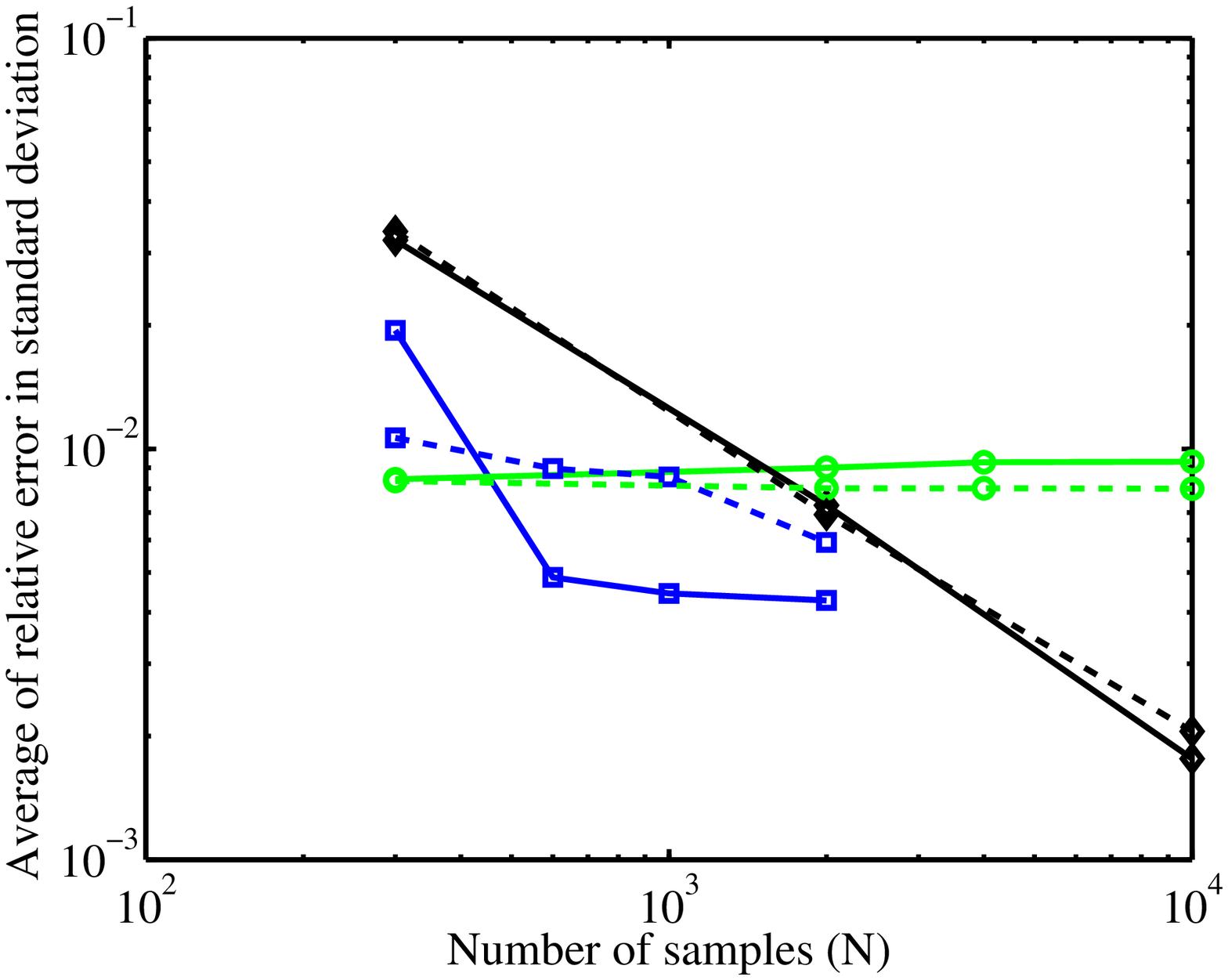}  
      &
      \includegraphics[width=2.7in]{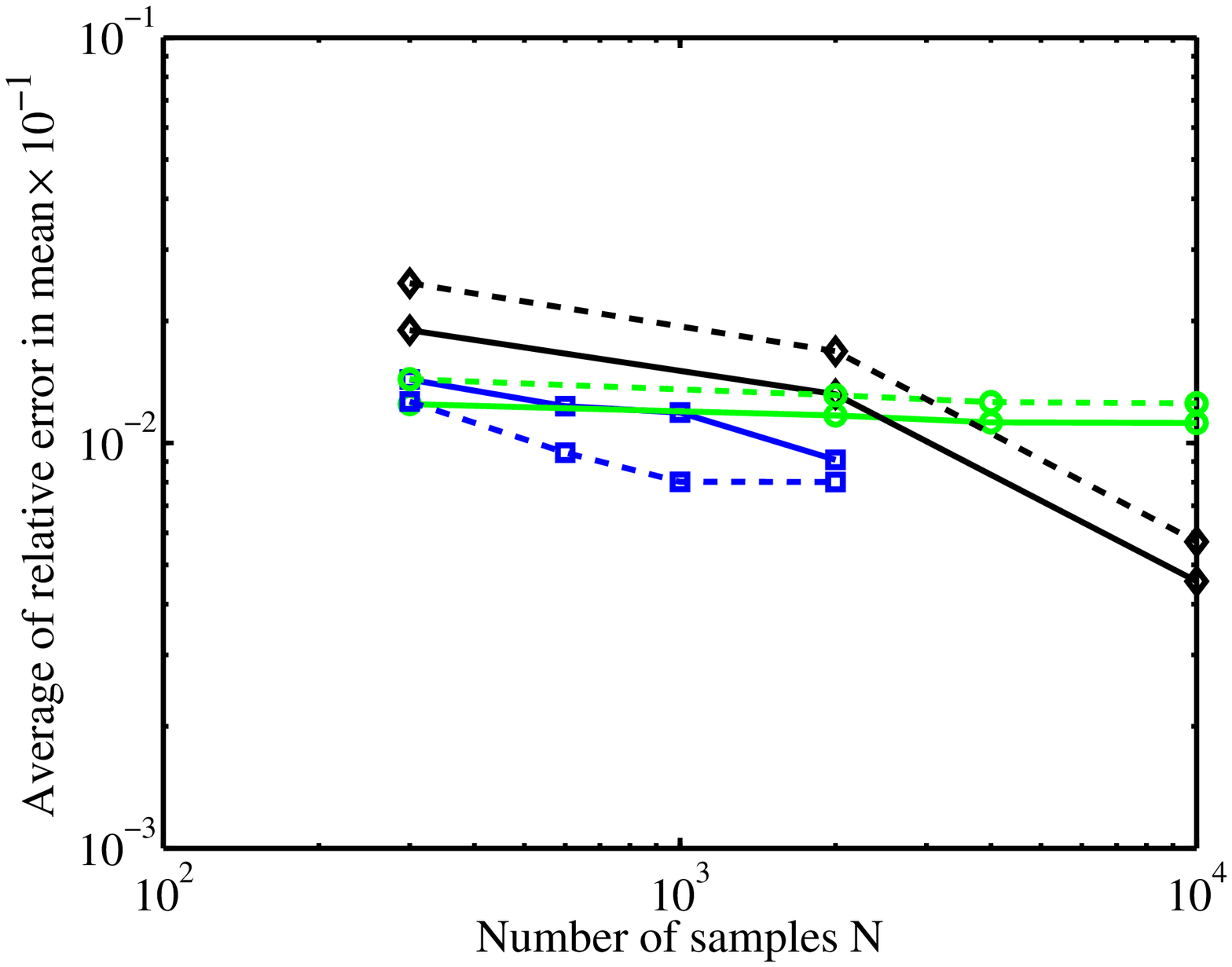} 
      \\
      (a) & (b)       
     \end{tabular}
      \caption{Comparison of the average of relative errors in standard deviation and mean for the separated representation, Regression, and Monte Carlo simulation (Line $\xi_1=0.5$ (\line(1,0){15}); Line $\xi_2=0.5$ ($---$)). (a) Average of relative error in standard deviation and (b) Average of relative error in  mean. (Separated representation ($\square$); PC regression ($\circ$); Monte Carlo ($\diamond$)).}            
\label{fig:Mean and STD errors for cavity flow}       
       \end{figure}

The solver is provided in FEniCS based on finite element algorithm and decoupling the governing equations and it ran $10,000$ times partially on the Janus supercomputer at UC Boulder. In figure \ref{fig:Cavity flow field}, the temperature contours for the 2D-cavity flow are shown, in which two vectors with size $n=20$ of temperature on vertical line $\xi_1=0.5$ and horizontal line $\xi_2=0.5$ are considered for separated representation approximation. The separated representation models were constructed following algorithm \ref{Algorithm:ALS}. Figure \ref{fig:Optimal rM for cavity flow} illustrates the optimal separated rank and polynomial degree for a set of numbers of samples $\left\lbrace 300,600,1000,2000\right\rbrace$. It can be observed that the separation ranks of the separated models are smaller than $10$, which lead to successfully approximate the function with \textit{low-rank} separated models. The separated representation complexities are sampling based and vary for different numbers of samples. Figures \ref{fig:Mean and STD errors for cavity flow}.a and \ref{fig:Mean and STD errors for cavity flow}.b compare the convergence of the average of standard deviation and mean of the scaled temperature on the two lines $\xi_1=0.5$ and $\xi_2=0.5$ obtained by separated representation, polynomial chaos expansion regression \citep{Doostan12}, and the standard Monte Carlo simulation. As the separated representation model construction is based on random sampling of the solution, the higher accuracy in the approximations may be achieved by incorporating more samples, while in the PC regression the solution accuracy may not improve by incorporating larger numbers of samples. Additionally, the convergence rate of the separated representation is faster than the PC regression. 

\begin{figure}
    \centering
    \begin{tabular}{cc}
            \hspace{-1cm}    
      \includegraphics[width=2.7in]{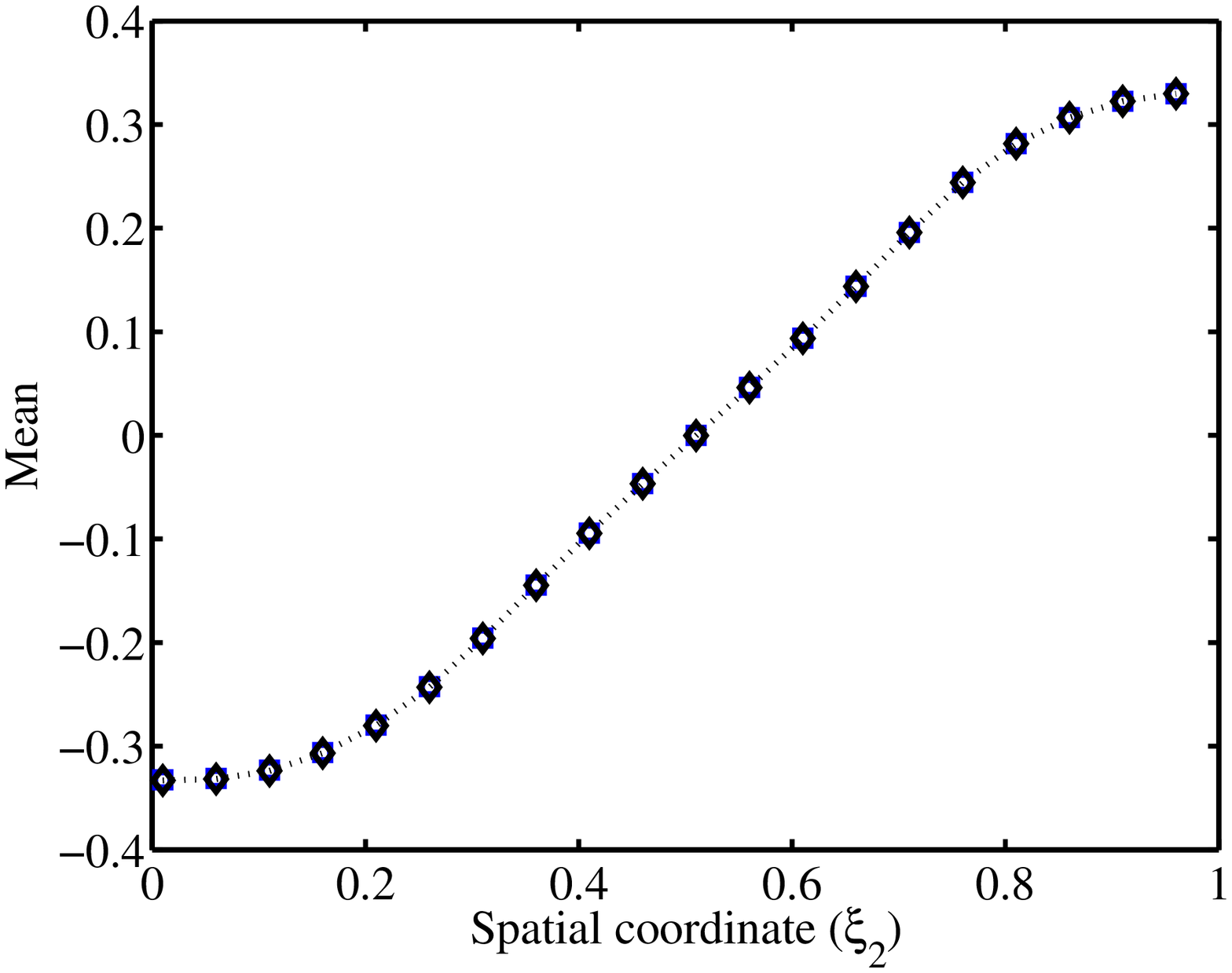}  
      &
      \hspace{-0.5cm}
      \includegraphics[width=2.7in]{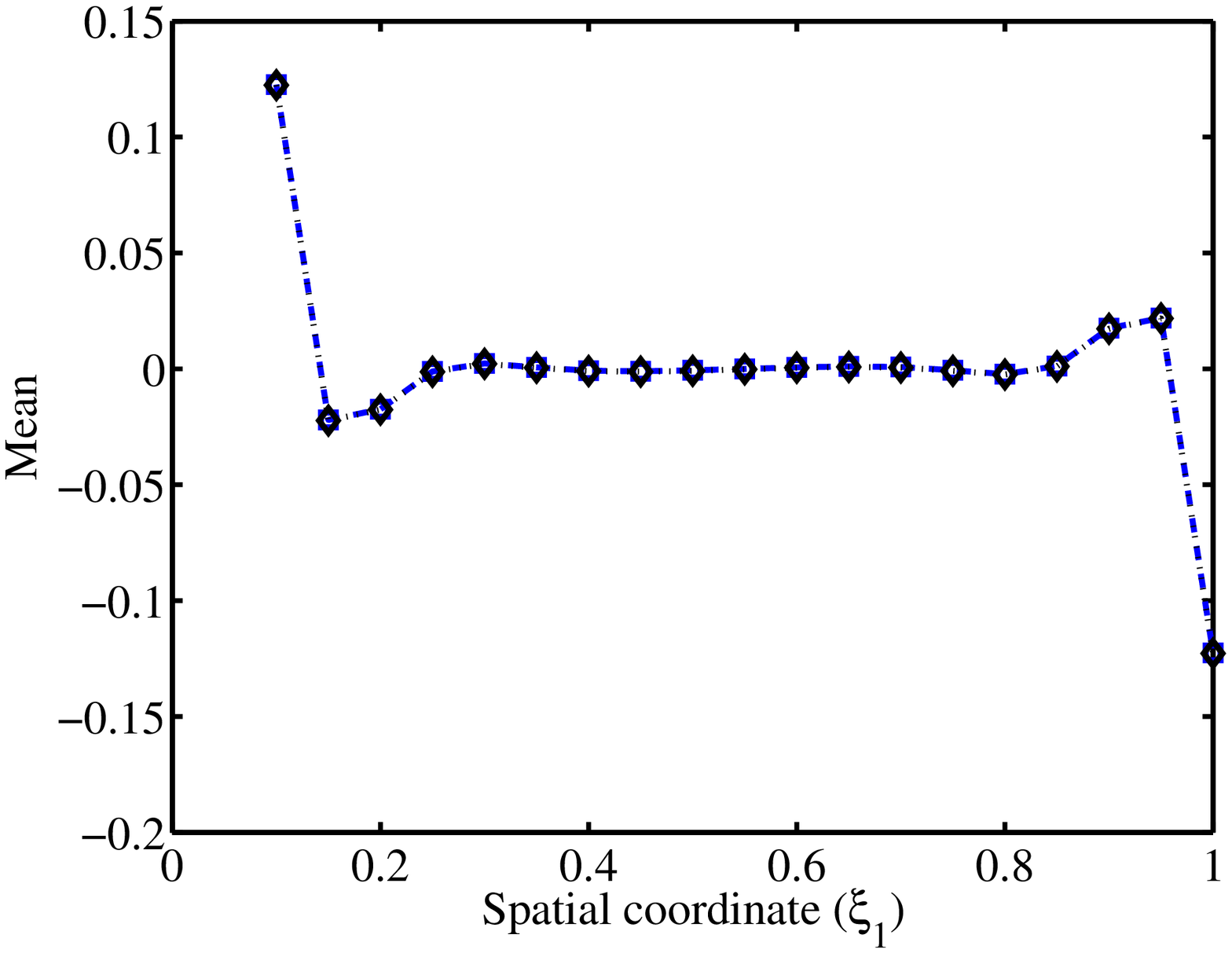} 
      \\
      (a) & (b)
      \\
      \hspace{-0.5cm}
      \includegraphics[width=2.7in]{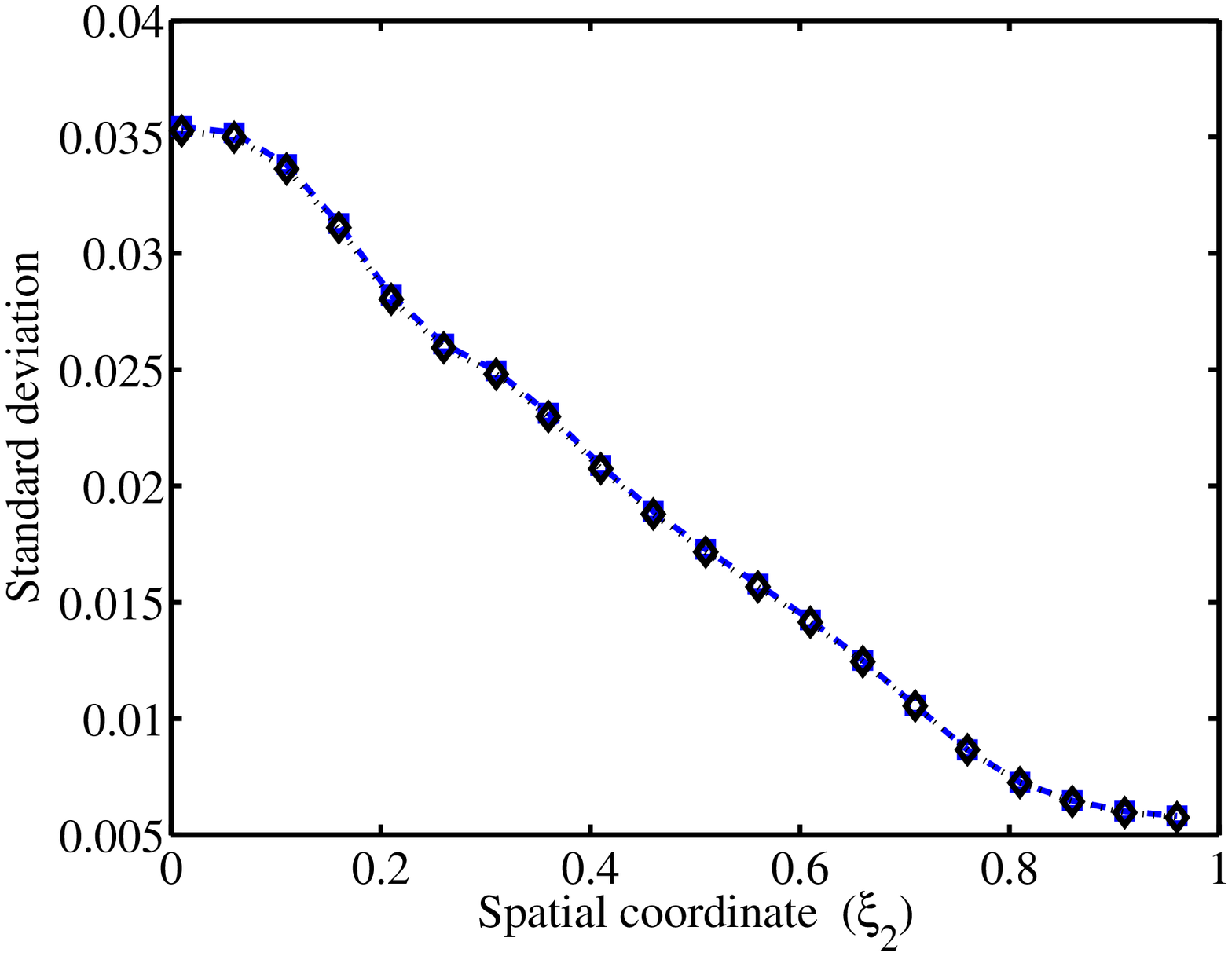}  
      &
      \hspace{-0.5cm}
      \includegraphics[width=2.7in]{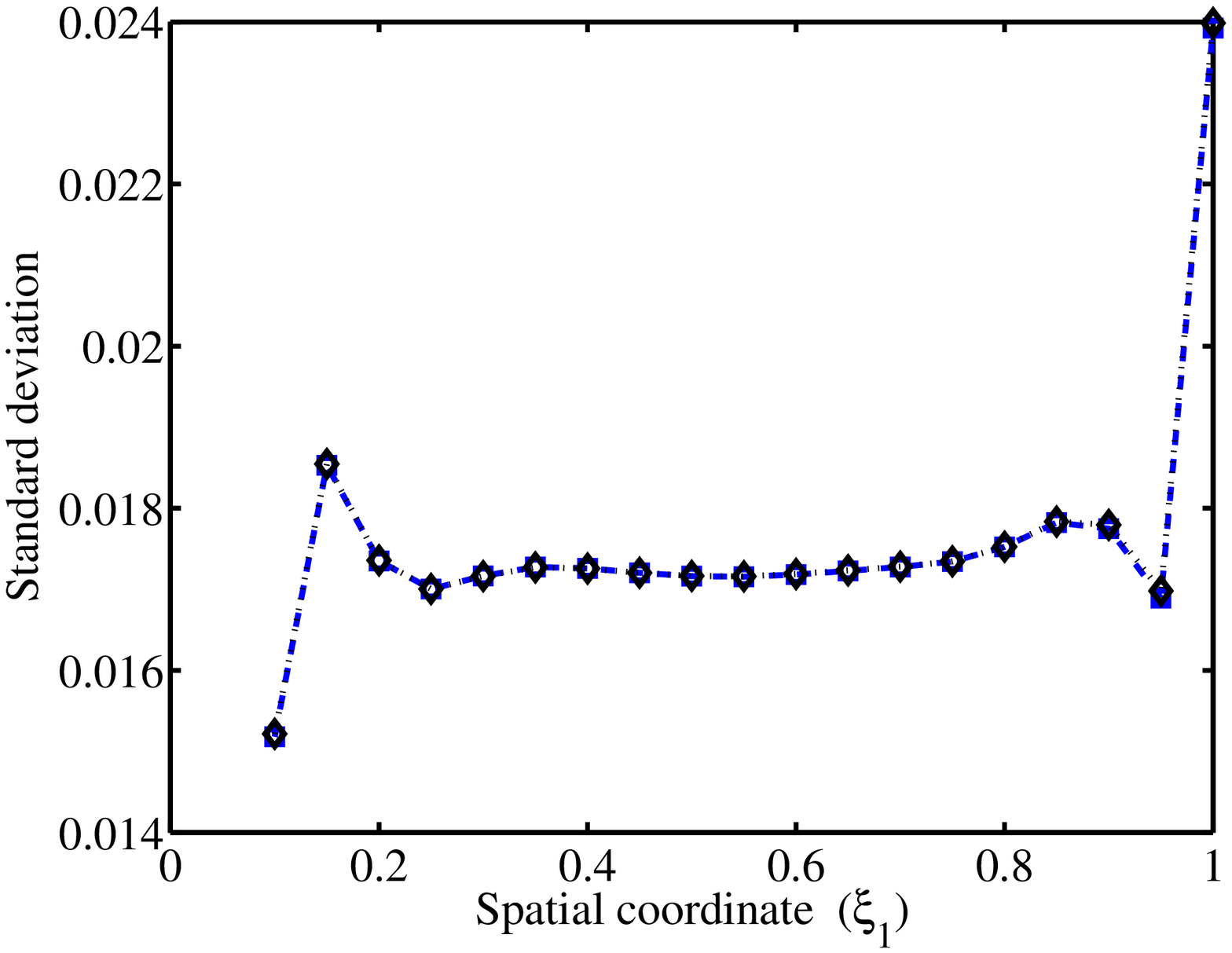}           
      \\
      (c) & (d)   
      
     \end{tabular}
      \caption{ The values of the mean and standard deviation of the scaled temperature as a function of space. (a) Mean value on line $\xi_1=0.5$; (b) Mean value on line $\xi_2=0.5$ (c) Standard deviation value on line $\xi_1=0.5$; (d) Standard deviation value on line $\xi_2=0.5$. (Estimated with Monte Carlo $(N=10,000)$ ({\scriptsize$\;\;\;\; \Diamond$} $\hspace{-.73cm} ---$) and Separated representation approximation $(N=2000)$  ($\;\;\;\; \square $ $\hspace{-.73cm} ---$)).}            
\label{fig:Mean and STD for two lines}       
       \end{figure}

To approximate the temperature and its standard deviation and mean values on both lines $\xi_1=0.5$ and $\xi_2=0.5$, the separated representation model which is constructed with $N=2000$ samples is used. Figure \ref{fig:Mean and STD for two lines} compares the approximations of the separated representation model with the standard Monte Carlo method. Figures \ref{fig:Mean and STD for two lines}.a and \ref{fig:Mean and STD for two lines}.b illustrate the mean values and figures \ref{fig:Mean and STD for two lines}.c and \ref{fig:Mean and STD for two lines}.d illustrate the standard deviation values. The average of relative errors in standard deviation and mean on the line $\xi_1=0.5$ are $\left\lbrace 0.008, 0.00059 \right\rbrace$ and corresponding values for the line $\xi_2=0.5$ are $\left\lbrace 0.009, 0.00042 \right\rbrace$, which show the out performance of the separated representation approximation.


\begin{figure}
    \centering
    \begin{tabular}{cc}
            \hspace{-0.5cm}    
      \includegraphics[width=2.6in]{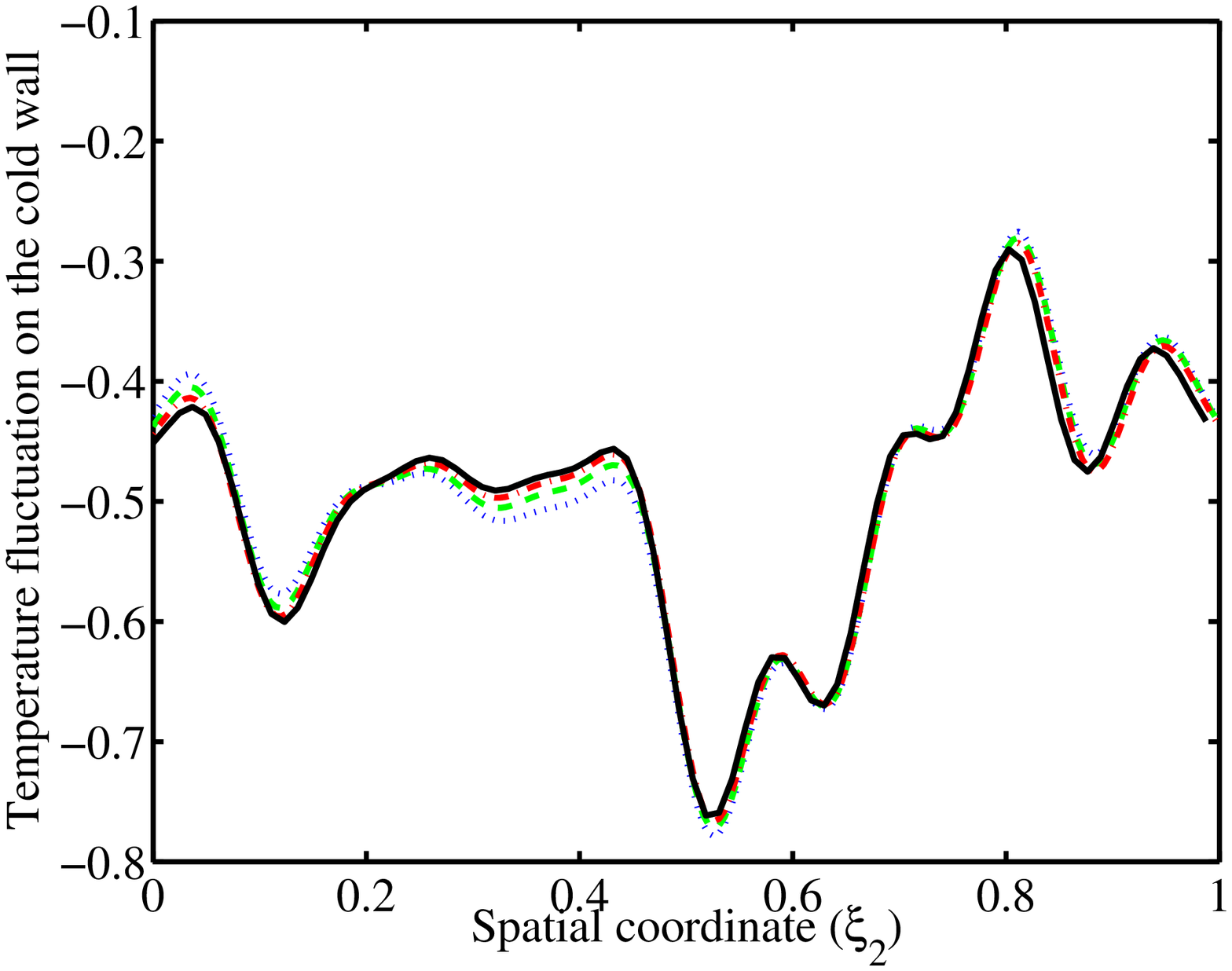}  
      &
      \includegraphics[width=2.7in]{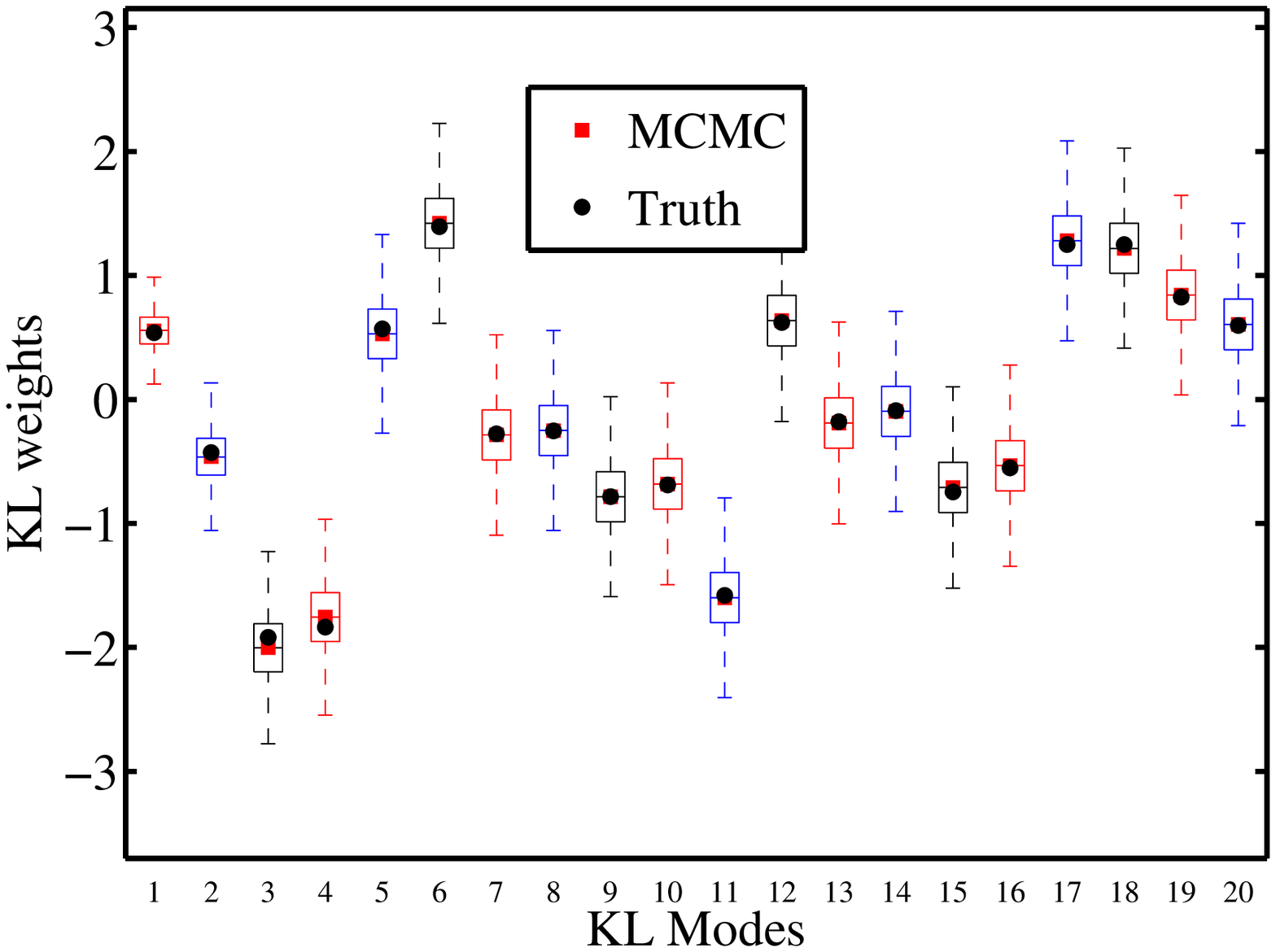} 
      \\
      (a) & (b)            
     \end{tabular}
      \caption{MCMC results for 2D cavity flow. (a) Temperature fluctuations on the cold wall obtained with posterior realizations vs. the spatial variable. ($\sigma_p = 0.3$ ($\;\;\;\;\;\;  $ $\hspace{-.69cm} .-.-.-$),  $\sigma_p = 0.6$ ($\;\;\;\;\; $ $\hspace{-.69cm} ---$),  $\sigma_p = 1$ ($\;\;\;\;\;\; $ $\hspace{-.69cm} \ldots$), and truth (solid line). (b) Posterior boxplot obtained with the MCMC. (Posterior mean {\scriptsize$\left( \square \right)$} and the true value of KL weights $\left(\circ\right)$.}            
\label{fig:MCMC results for cavity}       
       \end{figure}

\begin{figure}  
\centering
\begin{tabular}{c}
\hspace{-0.5cm}    
\includegraphics[width=5.5in]{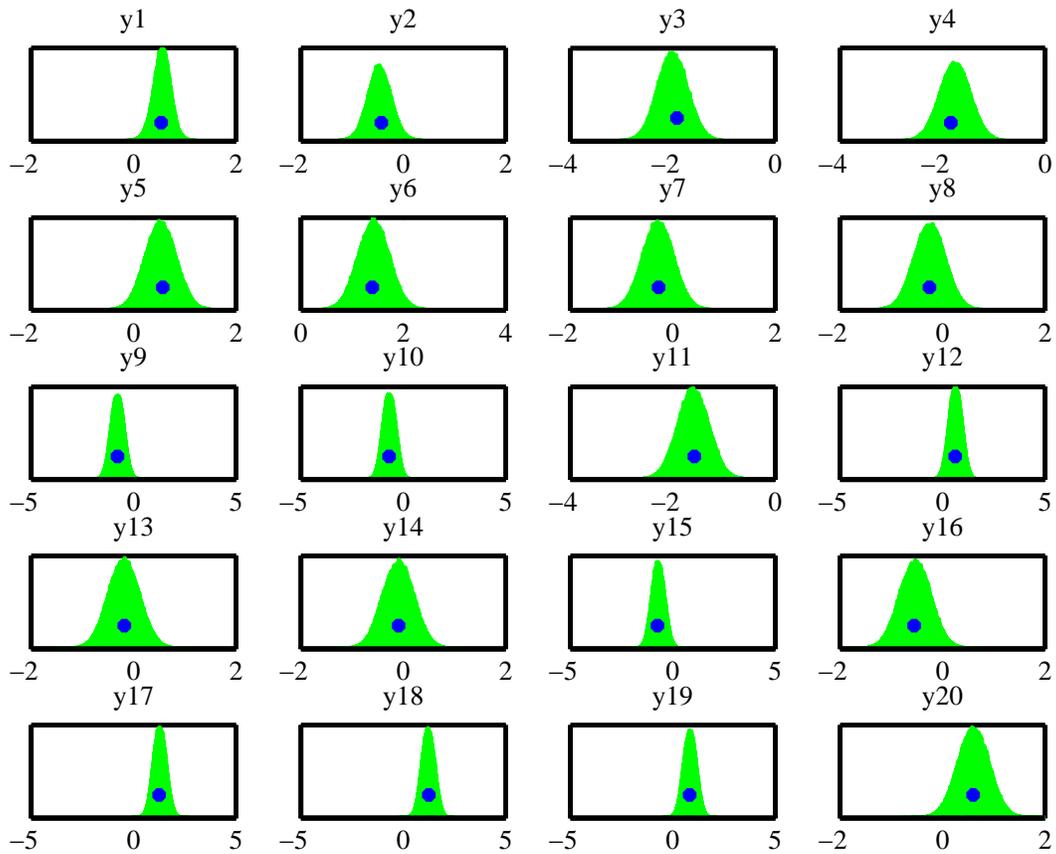}  
\end{tabular}
\caption{The histogram of MCMC chain for 2D cavity flow. (The true value of the realizations $\left( \circ \right)$)}            
\label{fig:histogram of y}       
\end{figure}

Here, the analysis is focused on the posterior distributions of KL mode weights in estimating the temperature fluctuations on the cold wall. In inverse modelling, Gaussian prior distributions with a prior mean computed by the method in \citep{Jeffrey98} and prior standard deviations $\left\lbrace 0.3,0.6,1\right\rbrace$ are assumed, where the results are shown in figure \ref{fig:MCMC results for cavity}.a. The solid line represents the exact value of the fluctuations used to generate data before adding the observational noise term, and the rest of the lines represent the fluctuations computed by eq (\ref{eq:cold wall temp fluctuation}) with the mean of the MCMC chains for $\left\lbrace y_i \right\rbrace_{i=1}^{d=20}$. The MCMC results with different prior standard deviations were obtained using a DRAM with $10^6$ samples, discarding the first $2.5\times10^5$ as burn-in samples. It can be seen that by decreasing the prior standard deviation, which restricts the sampling space around the prior mean and may lead to selecting more accurate samples, more features of the temperature fluctuations can be captured. In figure \ref{fig:MCMC results for cavity}.b a boxplot of the exact values of the KL mode weights is illustrated, superimposed with the posterior mean of the MCMC chain. It is shown that the mode weights from lower indices to higher indices are identified accurately, and the exact values and MCMC means agree reasonably well. For better illustration of the method performance, in figure \ref{fig:histogram of y}, the histograms of the MCMC chain for $\left\lbrace y_i \right\rbrace_{i=1}^{d=20}$ are plotted. In this figure the exact values of $y_i$ are superimposed on the histogram of the posterior samples. The density of variables generated by MCMC coincides with the exact value of random variables.  

Each MCMC chain takes approximately 24 hours with two 4-cores CPU and 12 GB RAM computer, however inverse modelling using the physical model generally deemed to be prohibitive. It is not claimed that this is the most efficient inverse modelling approach, however, it is claimed that the separated representation surrogate model not only provides opportunities to do inverse modelling, but also it represents an out performance MCMC approach.

\section{Conclusion}
\label{sec: Conclusion}

In this study, a surrogate model has been introduced non-intrusively in the context of the low-rank separated representation, which makes feasible the use of an intractable Markov Chain Monte Carlo (MCMC) simulation in Bayesian inference for high-dimensional stochastic functions. In the separated representation approach, a high-dimensional stochastic function is broken down into a linear sum of unknown one-dimensional functions of random inputs. Here, the separated model approximates a vector of a continuous solution at discrete values of a physical variable. This vector valued separated model, which is an extension of previous work \cite{Doostan12} for the scalar-valued case, leads to a significant reduction in the computational cost of the approximation by an order of magnitude equal to the vector size. Because the solution can be approximated by one vector valued separated model, while the solution at each physical variable must be approximated separately with different scalar-valued separated models. An alternative least square regression-based approach was presented to stably construct the separated models. Also, the issue of instability which may occur in regression-based approaches was tackled using the Tikhonov regularization. The regularization is applied with a roughening matrix computing the gradient of the solution, which leads to have more control over the solutions and penalizing and smoothing the higher order polynomials. In order to find an adequate regularization parameter, Generalized Cross Validation (GCV) is adopted. Furthermore, a perturbation-based error indicator has been defined to find the optimal model complexities known as separation rank and polynomial degree. These parameters are independent from the function dimensionality $d$, which might lead to a successful approximation with a number of randomly generated realizations of stochastic functions linearly depends on $d$. The computational cost of the approximation quadratically increases with respect to the function dimensionality which may overcome the issue of the curse of dimensionality, a bottle-neck for uncertainty quantification of high-dimensional stochastic functions.

 It has been shown numerically that the low-rank separated representation approximation model outperforms the current techniques for high-dimensional approximations. And also, using these surrogates particularly of high-dimensional stochastic functions, makes computationally prohibitive inverse problem analysis feasible with high accuracy. The performance of the approach was examined with three problems, including an 11-dimensional manufactured function, a 41-dimensional (1D in space) elliptic PDE, and a 21-dimensional (2D in space) cavity flow. Overall, the Bayesian inference with a surrogate separated representation model proceeds with more reliability and efficiency than with a physical model. 

For future works, the applications of low-rank separated representation approximations in sensitivity analysis and function dimensionality reduction might be attractive areas. 

\section*{Acknowledgements}

This work utilized the Janus supercomputer, which is supported by the National Science Foundation (award number CNS-0821794) and the University of Colorado Boulder. The Janus supercomputer is a joint effort of the University of Colorado Boulder, the University of Colorado Denver and the National Center for Atmospheric Research.

\bibliographystyle{plain}
\bibliography{significance and novelty of the paper}

\begin{thebibliography}{10}

\bibitem{Kolda09a}
E.~Acar, T.G. Kolda, and D.M. Dunlavy.
\newblock An optimization approach for fitting canonical tensor decompositions.
\newblock Technical report, Sandia National Laboratories, SAND2009-0857,
  Livermore CA, 2009.

\bibitem{Ammar06}
A.~Ammar, B.~Mokdad, F.~Chinesta, and R.~Keunings.
\newblock A new family of solvers for some classes of multidimensional partial
  differential equations encountered in kinetic theory modeling of complex
  fluids.
\newblock {\em Journal of Non-Newtonian Fluid Mechanics}, 139(3):153 -- 176,
  2006.

\bibitem{Aster05}
R.~C. Aster, B.~Borchers, and C.~H. Thurber.
\newblock {\em parameter estimation and inverse problems}.
\newblock Elsevier Academic Press, 2005.

\bibitem{Bayes1763}
T.~Bayes and R.~Price.
\newblock An essay towards solving a problem in the doctrine of chance. by the
  late rev. mr. bayes, communicated by mr. price, in a letter to john canton,
  m. a. and f. r. s.
\newblock Technical Report~53, Philosophical Transactions of the Royal Society
  of London, January 1763.

\bibitem{Beck09}
A.~Beck and M.~Teboulle.
\newblock A fast iterative shrinkage-threshold algorithm for linear inverse
  problems.
\newblock {\em SIAM J. Imaging Sciences}, 2:183--202, 2009.

\bibitem{Bernardo94}
J.~M. Bernardo and A.~F.~M. Smith.
\newblock {\em Bayesian Theory}.
\newblock Wiley, 1994.

\bibitem{Beylkin08}
G.~Beylkin, J.~Garcke, and M.~J. Mohlenkamp.
\newblock Multivariate regression and machine learning with sums of separable
  functions.
\newblock {\em SIAM Journal on Scientific Computing}, 31(3):1840--1857, 2009.

\bibitem{Beylkin02}
G.~Beylkin and M.J. Mohlenkamp.
\newblock {Numerical operator calculus in higher dimensions}.
\newblock {\em Proceedings of the National Academy of Science},
  99:10246--10251, 2002.

\bibitem{Beylkin05}
G.~Beylkin and M.J. Mohlenkamp.
\newblock Algorithms for numerical analysis in high dimensions.
\newblock {\em SIAM Journal on Scientific Computing}, 26(6):2133--2159, 2005.

\bibitem{Bjorck96}
A.~Björck.
\newblock {\em Numerical methods for least squares problems}.
\newblock 1996.

\bibitem{Bremaud99}
P.~Br{\'e}maud.
\newblock {\em Markov Chains, Gibbs Fields, Monte Carlo Simulation, and
  Queues}.
\newblock Springer, 1999.

\bibitem{Constantine11a}
P.~Constantine, Q.~Wang, A.~Doostan, and G.~Iaccarino.
\newblock A surrogate accelerated bayesian inverse analysis of the {HyShot II}
  flight data.
\newblock In {\em AIAA-2011-2037}, 2011.

\bibitem{Daubechies04}
I.~Daubechies, M.~Defrise, and C.~De Mol.
\newblock An iterative thresholding algorithm for linear inverse problems with
  a sparsity constraint.
\newblock {\em Communications on Pure and Applied Mathematics},
  57(11):1413--1457, 2004.

\bibitem{Doostan07}
A.~Doostan, R.~Ghanem, and J.~Red-Horse.
\newblock Stochastic model reduction for chaos representations.
\newblock {\em Computer Methods in Applied Mechanics and Engineering},
  196(37-40):3951--3966, 2007.

\bibitem{Doostan12}
A.~Doostan, A.~Validi, and G.~Iaccarino.
\newblock Non-intrusive low-rank separated approximation of high-dimensional
  stochastic models.
\newblock {\em Computer Methods in Applied Mechanics and Engineering}, 2013,
  doi: http://dx.doi.org/ 10.1016/j.cma.2013.04.003.

\bibitem{stephen06}
S.~E. Fienberg.
\newblock When did bayesian inference become ``bayesian"?
\newblock {\em Bayesian Analysis}, 1:1--40, 2006.

\bibitem{Furukawa02}
R.~Furukawa, H.~Kawasaki, K.~Ikeuchi, and M.~Sakauchi.
\newblock Appearance based object modeling using texture database: acquisition,
  compression and rendering.
\newblock In {\em EGRW 02: Proceedings of the 13th Eurographics workshop on
  Rendering, Aire-la-Ville, Switzerland, Switzerland}, pages 257--266, 2001.

\bibitem{Gelman03}
A.~Gelman, J.B. Carlin, H.S. Stern, and D.B. Rubin.
\newblock {\em Bayesian Data Analysis, 2nd Edition}.
\newblock Chapman {\&} Hall, Boca Raton, 2003.

\bibitem{Geweke89}
J.~Geweke.
\newblock Bayesian inference i n econometric models using monte carlo
  integration.
\newblock {\em Econometrica}, 57:1317--1339, 1989.

\bibitem{Geweke92}
J.~Geweke.
\newblock Bayesian inference i n econometric models using monte carlo
  integration.
\newblock {\em Econometrica}, 57:1317--1339, 1989.

\bibitem{Gilks96}
W.~R. Gilks, S.~Richardson, and D.~J. Spiegelhalter.
\newblock {\em {Markov chain Monte-Carlo in pratice}}.
\newblock 1996.

\bibitem{GreenP01}
P.~Green and A.~Mira.
\newblock Delayed rejection in reversible jump metropolis-hastings.
\newblock {\em Biometrika}, 88:1035--1053, 2001.

\bibitem{Charles93}
C.~Groetsch.
\newblock {\em {Inverse problems in the mathematical sciences}}.
\newblock 1993.

\bibitem{Haario04}
H.~Haario, M.~Laine, and E.~Saksman M.~Lehtinen, and J.~Tamminen.
\newblock Monte carlo methods for high dimensional inversion in remote sensing.
\newblock {\em J. R. Statist. Soc. B}, 66:591--608, 2004.

\bibitem{Heikki06}
H.~Haario, M.~Laine, A.~Mira, and E.~Saksman.
\newblock {DRAM: Efficient adaptive MCMC}.
\newblock {\em Statistics and Computing}, 16:339--354, 2006.

\bibitem{Haario99b}
H.~Haario, E.~Saksman, and J.~Tamminen.
\newblock Adaptive proposal distribution for random walk metropolis algorithm.
\newblock {\em Comp. Stat.}, 14:375--393, 1999.

\bibitem{Heikki98}
H.~Haario, E.~Saksman, and J.~Tamminen.
\newblock {An Adaptive Metropolis algorithm}.
\newblock {\em Bernoulli}, 7:223--242, 2001.

\bibitem{Hackbusch04}
W.~Hackbusch and B.~N. Khoromskij.
\newblock Kronecker tensor-product approximation to certain matrix-valued
  functions in higher dimensions.
\newblock Technical Report Preprint 16, Max-Planck-Institut {f\"ur} Mathematik
  in den Naturwissenschaften, 2004.

\bibitem{Hansen10}
P.~C. Hansen.
\newblock {\em Discrete Inverse Problems Insight and Algorithms}.
\newblock SIAM, 2010.

\bibitem{Dave}
D.~Higdon and C.~Holloman.
\newblock {Markov chain Monte Carlo-based approaches for inference in
  computationally intensive inverse problems}.

\bibitem{Hitchcock}
F.L. Hitchcock.
\newblock The expression of a tensor or a polyadic as a sum of products.
\newblock {\em Journal of Mathematics and Physics}, 6:164--189, 1927.

\bibitem{Jaakkola2000}
T.~Jaakkola and M.~Jordan.
\newblock Bayesian parameter estimation via variational methods.
\newblock {\em Statistics and Computing}, 10:25--37, 2000.

\bibitem{Kaipio04}
J.~Kaipio and E.~Somersalo.
\newblock {\em Statistical and Computational Inverse Problems}.
\newblock Springer, 2004.

\bibitem{Andreas96}
A.~Kirsch.
\newblock An introduction to the mathematical theory of inverse problems.
\newblock {\em Applied Mathematical Sciences}, 120, 96.

\bibitem{Kolda09b}
T.~G. Kolda and B.~W. Bader.
\newblock Tensor decompositions and applications.
\newblock {\em SIAM Review}, 51(3):455--500, 2009.

\bibitem{Kroonenberg80}
P.~Kroonenberg and J.~Leeuw.
\newblock Principal component analysis of three-mode data by means of
  alternating least squares algorithms.
\newblock {\em Psychometrika}, 45(1):69--97, March 1980.

\bibitem{Jeffrey98}
J.~C. Lagarias, J.~A. Reeds, M.~H. Wrights, and P.~E. Wright.
\newblock Convergence properties of the nelder-mead simplex method in low
  dimensions.
\newblock {\em SIAM J. Optim.}, 9:112--147, 1998.

\bibitem{Ma09b}
X.~Ma and N.~Zabaras.
\newblock An efficient bayesian inference approach to inverse problems based on
  an adaptive sparse grid collocation method.
\newblock {\em Inverse Problems}, 25:035013, 2009.

\bibitem{LeMaitre10}
O.P.~Le Maitre and O.~Knio.
\newblock {\em Spectral Methods for Uncertainty Quantification with
  Applications to Computational Fluid Dynamics}.
\newblock Springer, 2010.

\bibitem{Marzouk09a}
Y.~M. Marzouk and H.~N. Najm.
\newblock Dimensionality reduction and polynomial chaos acceleration of
  bayesian inference in inverse problems.
\newblock {\em J. Comput. Phys.}, 228:1862--1902, 2009.

\bibitem{Youssef09}
Y.~M. Marzouk and D.~Xiu.
\newblock {A Stochastic Collocation Approach to Bayesian Inference in Inverse
  Problems}.
\newblock {\em Communications in Computational Physics}, 6:826--847, 2009.

\bibitem{Marzouk09b}
Y.~M. Marzouk and D.~Xiu.
\newblock A stochastic collocation approach to bayesian inference in inverse
  problems.
\newblock {\em Communications In Computational Physics}, 6(4):826--847, 2009.

\bibitem{McKeague05}
I.~McKeagueand, G.~K. Nicholls, and K.~Speer.
\newblock Statistical inversion of south atlantic circulation in an abyssal
  neutral density layer.
\newblock {\em Journal of Marine Research}, 63:683--704, 2005.

\bibitem{Metropolis53}
N.~Metropolis, A.~W. Rosenbluth, M.~N. Rosenbluth, A.~H. Teller, and E.~Teller.
\newblock Equation of state calculations by fast computing machines.
\newblock {\em The Journal of Chemical Physics}, 21:1087--1092, 1953.

\bibitem{Miller04}
I.~Miller and M.~Miller.
\newblock {\em Mathematical Statistics with Application}.
\newblock Pearson, 2004.

\bibitem{Mira02}
A.~Mira.
\newblock Ordering and improving the performance of monte carlo markov chains.
\newblock {\em Bernoulli}, 16:340--350, 2002.

\bibitem{Morozov93}
V.~A. Morozov.
\newblock {\em Regularization Methods for Ill-Posed Problems}.
\newblock CRC, 1993.

\bibitem{Tarek12}
T.~Moselhy and Y.~M. Marzouk.
\newblock Bayesian inference with optimal maps.
\newblock {\em Arxiv:1109.1516v3}, 2012.

\bibitem{Rao99}
C.~R. Rao and H.~Toutenburg.
\newblock {\em Linear Models: Least Squares and Alternatives}.
\newblock 1999.

\bibitem{Shashua01}
A.~Shashua and A.~Levin.
\newblock Linear image coding for regression and classification using the
  tensor-rank principle.
\newblock In {\em CVPR 2001: Proceedings of the 2001 IEEE Computer Society
  Conference on Computer Vision and Pattern Recognition}, pages 42--49, 2001.

\bibitem{Smith84}
R.~L. Smith.
\newblock Efficient monte carlo procedures for generating points uniformly
  distributed over bounded regions.
\newblock {\em Oper. Res.}, 32:1296--1308, 1984.

\bibitem{Spall10}
J.~C. Spall.
\newblock {\em Introduction to Stochastic Search and Optimization, Estimation,
  Simulation and Control}.
\newblock Wiley, 2010.

\bibitem{Stewart79}
T.~Stewart.
\newblock Multiparameter univariate bayesian inference.
\newblock {\em J. Amer. Statist. Assoc.}, 74:684--693, 1979.

\bibitem{Tarantola05}
A.~Tarantola.
\newblock {\em {Inverse problem theory - and methods for model parameter
  estimation}}.
\newblock 2005.

\bibitem{Tierney94}
L.~Tierney.
\newblock Markov chains for exploring posterior distributions.
\newblock {\em The Annals of Statistics}, 22(4):1701--1762, 1994.

\bibitem{Tropp10a}
J.A. Tropp and S.J. Wright.
\newblock Computational methods for sparse solution of linear inverse problems.
\newblock {\em Proceedings of the IEEE}, 2010.
\newblock in press.

\bibitem{Vogel02}
C.~R. Vogel.
\newblock {\em Computational Methods for Inverse Problems}.
\newblock Society for Industrial and Applied Mathematics, Philadelphia, PA,
  USA, 2002.

\bibitem{Wan05}
X.~Wan and G.~Karniadakis.
\newblock An adaptive multi-element generalized polynomial chaos method for
  stochastic differential equations.
\newblock {\em J. Comp. Phys.}, 209:617--642, 2005.

\bibitem{Zellner84}
Zellner and E.~Rossi.
\newblock Bayesian analysis of dichotomous quantal response models.
\newblock {\em J. Econometrics}, 25:365--393, 1984.

\end{thebibliography}

\end{document}